\documentclass[acmsmall]{acmart}
\setcopyright{none}
\usepackage[utf8]{inputenc}

\usepackage{natbib}
\usepackage{listings}% http://ctan.org/pkg/listings
\usepackage{url}

\usepackage{microtype}

\usepackage{graphicx}
\usepackage{amsmath}
\usepackage{siunitx}
\usepackage{color}

\usepackage[nounderscore]{syntax}
\usepackage{mathpartir}

\begin{document}

\setlength{\pdfpageheight}{\paperheight}
\setlength{\pdfpagewidth}{\paperwidth}

% \newcommand{\secref}[1]{Sec.~\ref{sec:#1}}  % for use in text
% \newcommand{\Secref}[1]{Sec.~\ref{sec:#1}}  % for start of sentence
% \newcommand{\figref}[1]{Fig.~\ref{fig:#1}}     % for use in text
% \newcommand{\Figref}[1]{Figure~\ref{fig:#1}}   % for start of sentence
% \newcommand{\Figrefs}[2]{Figures~\ref{fig:#1} and~\ref{fig:#2}}
% \newcommand{\figrefsMany}[1]{Figs.~\ref{fig:#1}}
% %\newcommand{\lineref}[1]{Line \ref{lst:#1}}  % for use in text
% \newcommand{\linerefs}[2]{Lines \ref{lst:#1} and~\ref{lst:#2}}  % for use in text
% \newcommand{\appref}[1]{Appendix~\ref{ap:#1}}
% \newcommand{\tabref}[1]{Table~\ref{tab:#1}}  % for use in text
% \newcommand{\Tabref}[1]{Table~\ref{tab:#1}}  % for start of sentence

% \newcommand{\ras}[1]{{\color{blue} Ras: [{#1}]}}
% \newcommand{\kartik}[1]{{\color{blue} Kartik: [{#1}]}}

\author{Kartik Chandra}
% \affiliation{
%     \institution{Henry M. Gunn High School}
%     \city{Palo Alto}
% }
\affiliation{
    \institution{Stanford University}
    \city{Palo Alto}
}
% \email{kartikchandra@acm.org}

\author{Rastislav Bodik}
\affiliation{
    \institution{University of Washington}
    \city{Seattle}
}
% \email{bodik@cs.washington.edu}

\settopmatter{printacmref=false}
\setcopyright{none}
\renewcommand\footnotetextcopyrightpermission[1]{}
% \pagestyle{plain}
%\conferenceinfo{xxx}{July, 2017}
\copyrightyear{2017}
%\copyrightdata{978-1-nnnn-nnnn-n/yy/mm}
%\copyrightdoi{nnnnnnn.nnnnnnn}

\title{\toolname: Synthesis-Based Reasoning for Type Systems}

\newcommand{\toolname}{\textls[-50]{\textsc{Bonsai}}}
\newcommand{\nocaptionrule}[0]{}

\begin{abstract}

We describe algorithms for symbolic reasoning about executable models of type systems, supporting three queries intended for designers of type systems.  
First, we check for type soundness bugs and synthesize a counterexample program if such a bug is found. Second, we compare two versions of a type system, synthesizing a program accepted by one but rejected by the other. Third, we minimize the size of synthesized counterexample programs.

These algorithms symbolically evaluate typecheckers and interpreters, producing formulas that characterize the set of programs that fail or succeed in the typechecker and the interpreter.   
However, symbolically evaluating interpreters poses efficiency challenges, which are caused by having to merge execution paths of the various possible input programs. Our main contribution is the \emph{Bonsai tree}, a novel symbolic representation of programs and program states which addresses these challenges. Bonsai trees encode complex syntactic information in terms of logical constraints, enabling more efficient merging.

We implement these algorithms in the \toolname\ tool, an assistant for type system designers. We perform case studies on how \toolname\ helps test and explore a variety of type systems. \toolname\ efficiently synthesizes counterexamples for soundness bugs that have been inaccessible to automatic  tools, and is the first automated tool to find a counterexample for the recently discovered Scala soundness bug SI-9633~\cite{amintatepaper}.
\end{abstract}

\maketitle

\section{Introduction}

Today's type system designers strive to develop typecheckers that balance expressiveness and convenience with powerful static guarantees. This has led to a variety of innovations, such as polymorphism, path-dependent types, and ownership types. On their own, these features promise programmers strong static guarantees on their programs.
Unfortunately, \emph{combining} such features often creates intricate soundness bugs, some of which have gone unnoticed for many years. For example, soundness bugs were caused by the confluence of assignment and polymorphism in ML~\cite{mlbug1,mlbug2}, and of path-dependent types and nullable values in Scala~\cite{amintatepaper}.

To automate checking of type systems, we present a set of \emph{symbolic} algorithms that aid type system designers in reasoning about executable language models. Using bounded program synthesis techniques, we demonstrate how users can compute answers to queries such as these:
\begin{itemize}
\item \emph{Soundness.} If the type system is not sound, synthesize a counterexample, a program that passes the typechecker but fails in the interpreter.
\item \emph{Comparison.} Given two versions of a typechecker, synthesize a program accepted by one version but rejected by the other, elucidating the impact of changes to a type system.
\item \emph{Minimization.} For either of the above queries, produce the \emph{smallest} possible counterexample.
\end{itemize}

\begin{figure*}[bt]
\includegraphics[width=0.8\columnwidth]{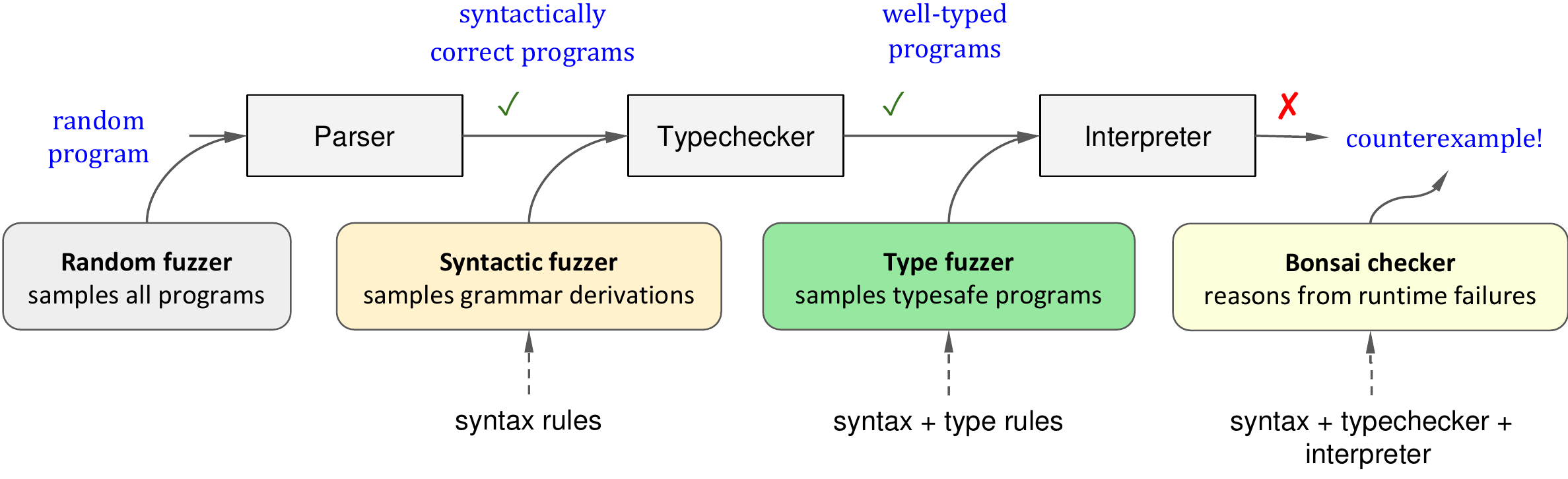}
\nocaptionrule
\caption{Searching for a counterexample using various combinations of goal-directed search and forward execution. For example, the \emph{type fuzzer} composes type judgments with the goal of generating a random type-safe program $P$. It then executes $P$ forward on the interpreter, hoping that $P$ will fail.  }
\label{fig:pipeline}
\end{figure*}

Successfully answering these queries requires us to efficiently explore the extremely large space of candidate programs that are complex enough to demonstrate a harmful interaction. Fuzzers, which are commonly used to answer the first query, attempt to search these large spaces by avoiding programs that would be rejected somewhere along the parser-typechecker-interpreter pipeline~(see~Figure~\ref{fig:pipeline}). Early fuzzers explored the space of \emph{all} programs, some of which may have failed in the parser~\cite{DBLP:conf/esec/Zeller99}. More recent ``syntax fuzzers'' such as  Redex~\cite{DBLP:conf/popl/KleinCDEFFMRTF12} generate only syntactically correct programs, thus shrinking the candidate space. Further advancements allow us to generate only type-safe programs~\cite{DBLP:conf/esop/FetscherCPHF15,pipal,DBLP:conf/kbse/DeweyRH14,rustfuzzing}. These ``type fuzzers'' use constraint solvers to search for a typesafe program in a goal-directed fashion; only the interpreter is executed forward. However, even among type-safe programs, counterexamples are scarce.

The natural next step, then, is to reason about the \emph{entire} pipeline in a goal-directed fashion, so that a counterexample can be directly constructed by the constraint solver by reasoning backwards from the possible failure points in the interpreter. We develop such goal-directed algorithm by symbolically compiling the entire parser-typechecker-interpreter pipeline into a single logical formula. This reduction also allows us to make the other queries mentioned above by modifying the structure of the formula.

Our contributions are: 
\begin{itemize}
\item \textbf{The Bonsai tree} (Sections~\ref{sec:motivation} and~\ref{sec:bonsai-trees}):
    We describe Bonsai trees, a symbolic representation of a bounded space of input programs and tree-shaped program states.  Bonsai trees enable symbolic evaluation of language models into logical formulas. Symbolic evaluation can be performed on standard trees, of course, but Bonsai-powered symbolic evaluation is more efficient and more effective, generating formulas that are smaller and easier to solve. 
    Bonsai trees reduce formula sizes by encoding the syntactic information in logical constraints rather than in the shape of the data structure.

We demonstrate several interesting properties of Bonsai trees (Sections~\ref{sec:bonsai-trees} and~\ref{sec:implementation}). 
First, Bonsai trees make it easier to avoid symbolically evaluating the parser-typechecker-interpreter pipeline as a monolithic composition of the three components.  Instead, we evaluate them separately, producing three simpler formulas that in turn represent syntactically correct programs; programs that pass the typechecker even though they may not be syntactically correct; and programs that succeed (or fail) in the interpreter even though they may syntactically incorrect or not typesafe.  These three sets, illustrated in Figure~\ref{fig:venn-diagram}, are then intersected by conjuncting the formulas, to obtain the set of counterexamples.  

\begin{figure*}[t]
 \includegraphics[width=0.8\columnwidth]{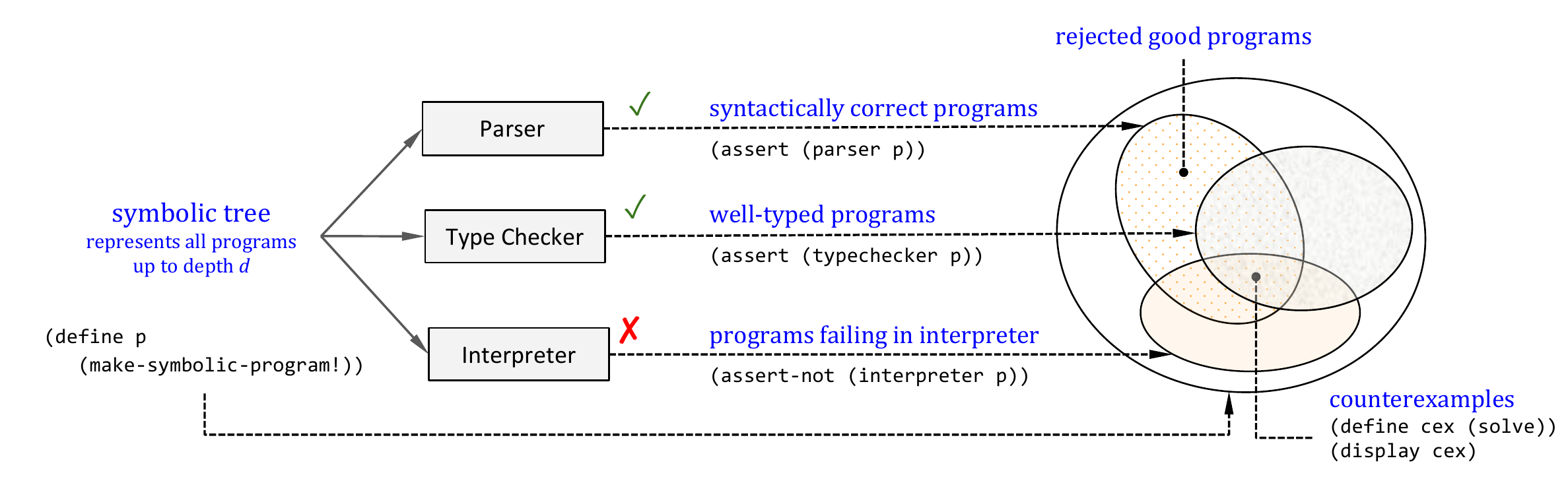}
 \nocaptionrule
 \caption{Bonsai performs three independent symbolic evaluations, executing the interpreter on trees that are both syntactically and type-incorrect. }
 \label{fig:venn-diagram}
\end{figure*}

Second, depending on what sets we intersect, we can formulate other queries.  For example, synthesis of correct programs that are rejected by the typechecker is also shown in Figure~\ref{fig:syntax-tree-repr}.  The \emph{Compare} query is formed analogously (see the bullet ``Case studies'' below for more details).  The \emph{Minimize} query simply adds the requirement that the solver returns a solution that minimizes some user-supplied metric, such as the tree size.

Third, Bonsai trees can be added to standard symbolic evaluators.  In particular, we formalize symbolic evaluation of Bonsai trees on a small core language (Section~\ref{sec:bonsai-trees}) and outline how we added this functional core language to Rosette~\cite{rosette}, a symbolic evaluator for a subset of Racket with assignments, objects, and other features (Section~\ref{sec:implementation}).  
This integration of Bonsai symbolic evaluation into Rosette is at the heart of the \toolname\ tool.

\item \textbf{Case studies} (Section~\ref{sec:casestudies}): 
We study a variety of type systems with \toolname, gauging its utility during a type system design. We perform three case studies, with Featherweight Java~\cite{featherweightjava}, Ownership Java~\cite{ownershipjava}, and DOT, the dependently-typed model of Scala~\cite{dotlanguage}.

We first examine how counterexamples can provide a clear explanation of soundness bugs. We demonstrate that by understanding how a counterexample fails in the interpreter, in our case by examining a stack trace of a program crash, we can better understand the cause of a soundness bug.

Second, we investigate what language specification patterns are friendly to symbolic evaluation. Interestingly, we find that Bonsai trees can be efficiently used to represent not only ASTs, but also other auxiliary data structures such as environments and class tables.
 
Third, we experiment with the \emph{Compare} query: we ask \toolname\ to explain a static type rule that appeared to us unnecessary (subsumed by other checks) or unnecessarily restrictive (rejecting correct programs).  We removed the rule, creating a more permissive version of the type system and asked \toolname\ ``Is there a program (i)~rejected by the old type system, (ii)~accepted by the new type system, and (iii)~that nevertheless runs without error?'' \toolname\ synthesized such a program, suggesting that it may be useful to design a less restrictive type system.
 
Finally, we use the DOT model to stress \toolname's expressiveness.  DOT is a rich calculus with dependent types, function types, records, intersection types, and recursive types.  We find DOT easy to reformulate to \toolname\ patterns.  We were able to express all features of DOT except for recursive types: early on we decided to use call-by-value semantics, only to realize later that DOT's definition of recursive types is only compatible with call-by-name semantics.  Overall, implementing DOT features did not hamper symbolic evaluation. In fact, in minutes, we synthesize on DOT a counterexample for the Scala soundness error SI-9633~\cite{amintatepaper}. 

\item \textbf{Performance evaluation} (Section~\ref{sec:performance}):
First, in under an hour, we synthesize a counterexample for the language with assignments and polymorphic references~\cite{DBLP:journals/iandc/Tofte90}, which has been inaccessible to automatic tools~\cite{DBLP:conf/esop/FetscherCPHF15}.  
Second, we show that Bonsai trees allow exploration of vastly larger spaces than the standard symbolic tree.  In our experiments with lambda calculus, \toolname\ explores $10^{50}$ vs. $10^{25}$ candidate programs in 20 seconds. 
Third, on a subset of Redex benchmarks, \toolname\ is about 600-times faster than a syntactic fuzzer and 12-times faster than a type fuzzer.  We also find \toolname\  superior in the size of counterexamples, synthesizing counterexamples about 10-times smaller, on average, than the type fuzzer.

\end{itemize}
Implementations of many of the case studies and evaluations can be found at \url{https://bitbucket.org/bonsai-checker/}.

Overall, we believe that \toolname\ complements the theorem-prover-based mechanization of a soundness proof, especially during design space exploration.  \toolname's strengths are automation and queries that go beyond soundness checking.  Among weaknesses is the inability to guarantee absence of bugs.  This limitation is somewhat compensated by the \toolname's efficiency: we observe that during an overnight run, it exhaustively explores candidates 2-~to 4-times larger than the counterexamples constructed by human experts (Table~\ref{tab:langs}), providing a safety margin.  The relationship with other tools is given in Section~\ref{sec:relatedwork}.

\section{Background and Motivation}
\label{sec:motivation}

Figure~\ref{fig:argrammar} shows a simple executable specification of the simply-typed arithmetic language~\cite{piercebook}. This particular specification has an unsound typechecker, due to an omitted type rule---it does not verify that the two branches of an ``if" term have the same type. We would like to automatically synthesize a counterexample to soundness for this language by symbolically translating the typechecker and interpreter to SMT formulas, and then querying a solver for a program that passes ``check" but fails in ``execute." In this section, we explain the scalability challenges presented by this strategy, motivating a new symbolic representation of abstract syntax trees. In the following section, we address those challenges by designing such a symbolic representation, the \emph{Bonsai Tree}. Finally, equipped with the Bonsai Tree, we return to the example above and demonstrate how to synthesize a counterexample.

\begin{figure}
\begin{tabular}{p{7cm}p{7cm}}
%\begin{quote}
\footnotesize
\begin{verbatim}
; syntax
(define arithmetic-syntax  
  '([exp zero (succ exp) (if exp exp exp) (zero? exp)
    ]))
\end{verbatim}
\begin{verbatim}
; evaluator
(define (execute t)  
  (tree-match t
    'zero (lambda () 0)

    '(succ _) (lambda (x) (+ (execute x) 1))

    '(if _ _ _) (lambda (c t f)
                  (if (execute c) 
                      (execute t) 
                      (execute f)))
      
    '(zero? _) (lambda (x) (= 0 (execute x)))))
\end{verbatim}
%\end{quote}
&
%\begin{quote}
\footnotesize
\begin{verbatim}
; type rules
(define (check t)  
  (tree-match t
    'zero (lambda () 'nat)

    '(succ _) (lambda (x)
                (assert  (eq? (check x) 'nat))
                'nat)

    '(if _ _ _) (lambda (c t f)
                  (assert  (eq? (check c) 'bool))
                  (define t+ (check t))
                  (define f+ (check f))
                 ;; (assert (eq? t+ f+))
                 ;; omitted!
                  t+)

    '(zero? _) (lambda (x)
                 (assert (eq? (check x) 'nat))
                 'bool)))
\end{verbatim}
%\end{quote}
\end{tabular}
\caption{An executable specification of the arithmetic language~\cite{piercebook} that we would like to check for soundness. \texttt{tree-match} is a pattern-matching macro, which checks an input AST against the given patterns, and, upon finding a match, executes the accompanying lambda with the contents of the blanks as arguments.}
\label{fig:argrammar}
\end{figure}

We begin with a brief overview of symbolic evaluation techniques, focusing in particular on the ``branch-and-merge" strategy for evaluating conditional statements. We then explain how currently-used symbolic data structures like ASTs are created by representing ``symbolic unions" of concrete trees, and why this structure makes branch-and-merge operations prohibitively expensive. Finally, we introduce the key idea behind our new tree encoding, ``Bonsai," which allows the symbolic evaluator to efficiently branch and merge, dramatically decreasing the time needed to create the SMT formula.

\subsection{Background: Symbolic Evaluation}
\label{sec:symex}

Symbolic evaluation was first described by King~\cite{king-76}: the goal is to translate a program $P$ with inputs $x_1, x_2, \dots, x_n$ into an equivalent SMT formula $f(x_1, x_2, \dots, x_n)$, which can then be processed using an SMT solver. The inputs $x_i$ are called \emph{symbolic constants}; they can be booleans, integers, bitvectors, or members of any other theory supported by the SMT solver.
Importantly, substituting a concrete value $v_i$ for each $x_i$ in the formula should simplify to the same result as evaluating $P$ with arguments $v_i$ directly. For example, the symbolic evaluation of the program
\begin{quote}
\footnotesize
\begin{verbatim}
P(x1 : int) : int {
    int k = 0;
    for (int i=0; i<3; i++) { k += x1; }
    return k;
}
\end{verbatim}
\end{quote}
yields the formula $(((0+x_1)+x_1)+x_1)$.

A program may also make \emph{assertions} about its symbolic constants. For example, the statement \texttt{assert P(x1) = 9} would create the formula $(((0+x_1)+x_1)+x_1) = 9$. Such an assertion can then be sent as part of a query to the SMT solver, which would report the solution model $\{x_1 = 3\}$. Symbolic evaluators generally collect all assertions encountered during the program's execution in a set called the \emph{assertion store}. The final query is the conjunction of all assertions in the store.

The symbolic evaluation of most terms is a straightforward application of partial evaluation. However, branching statements require some care. Consider the program below.
\begin{quote}
\footnotesize
\begin{verbatim}
P(x1 : int) : string {
    if (x1 > 5) { return "big"; }
    else        { return "small"; }
}
\end{verbatim}
\end{quote}
Notice that the value of the condition \texttt{x1 > 5} is not known to the symbolic evaluator. Thus, the symbolic evaluator must \emph{branch} to try both paths, and then \emph{merge} the results to create a formula expressing both alternatives. Such a formula is known as an \emph{ite} for \emph{if-then-else}, and the entire process is known as a ``branch-and-merge operation." Here, the branch-and-merge operation yields $\mathsf{ite}(x_1>5, \texttt{"big"}, \texttt{"small"})$. The conditions $x_1>5$ and $\neg(x_1>5)$ are called the \emph{path conditions} of their respective values \texttt{"big"} and \texttt{"small"}.

Operations on an ite are mapped over its branches. For example, the program \texttt{strLen P(x1)} produces $\mathsf{ite}(x_1>5, \texttt{strLen "big"}, \texttt{strLen "small"})$, which is then simplified to $\mathsf{ite}(x_1>5, 3, 5)$. Thus, one could query the solver for $x_1$ such that \texttt{assert strLen P(x1) = 5}. The solver may report `2' as a solution. Following \citet{rosette}, we generalize the ite to the \emph{symbolic union}, which represents a sequence of $n$ chained ites. We represent $\mathsf{ite}(\phi_1, v_1, \mathsf{ite}(\phi_2, v_2, \dots))$ as the symbolic union $\{\phi_1 \rightarrow v_1, \phi_2 \rightarrow v_2, \dots, \phi_n \rightarrow v_n\}$. Each $\phi_i$ is a path condition to its corresponding value $v_i$; as with ites, operations on symbolic unions are mapped over each such value.

One final nuance lies in assertions made in the bodies of conditionals. Consider the program
\begin{quote}
\footnotesize
\begin{verbatim}
P(x1 : int) : void {
    if (x1 > 5) { assert x1 = 10; } // A
    else        { assert x1 =  0; } // B
}
\end{verbatim}
\end{quote}
A naive execution of this program creates the unsatisfiable assertion store $x_1 = 10 \wedge x_1 = 0$. However, both $10$ and $0$ should be valid values for $x_1$. To allow this, the symbolic evaluator maintains a \emph{global} path condition, which reflects the conditions under which an assertion is made. The path condition is updated every time the program branches with a symbolic condition. For example, when the first assertion (``A") is made, $x_1 > 5$ must hold. Thus, the path condition is $x_1 > 5$, and the actual assertion recorded by the symbolic evaluator is the implication $x_1>5 \implies (x_1 = 10)$. Similarly, the second assertion (``B") is recorded as $\neg(x_1>5)\implies (x_1=0)$. Thus, the query to the solver becomes $(x_1>5 \implies x_1 = 10) \wedge (\neg(x_1>5)\implies x_1=0)$, which is satisfiable with both $10$ and $0$ as solutions for $x_1$.

\subsection{Symbolic Syntax Trees}

To symbolically evaluate typecheckers and interpreters, we must represent programs in terms of a set of symbolic constants. The classical approach to representing symbolic ASTs uses symbolic unions to represent choosing between production rules. For the remainder of this section, we will consider a simple grammar for the lambda calculus, with the production rules $e \rightarrow x \;\vert\; \lambda x . e \;\vert\; (e\;e)$, where $x$ ranges over variable names.

\paragraph{Structure}

\begin{figure}
\includegraphics[width=0.5\columnwidth]{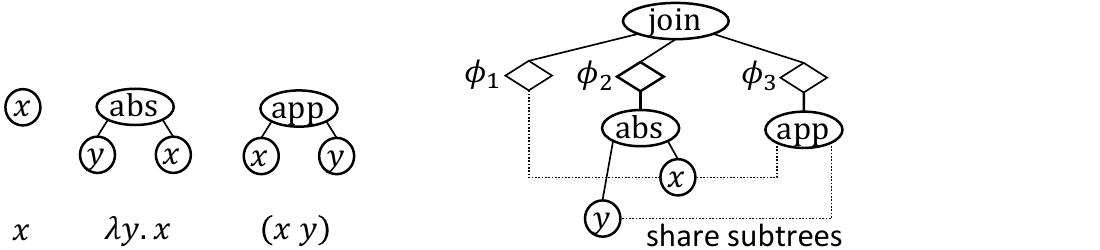}
\caption{The symbolic union of three syntax trees $G[x]$, $G[\lambda y.x]$, and $G[(x \; y)]$ under their respective path conditions $\phi_1$, $\phi_2$, and $\phi_3$. The ``join" node represents a symbolic union. Notice that since members of a symbolic union are disjoint, we can optimize for space by sharing subtrees across the merge.}
\label{fig:syntax-tree-repr}
\end{figure}

To construct a symbolic AST, one might begin by creating a symbolic union over a symbolic constant $c$, which chooses among the three types of depth-1 trees, i.e. the three production rules: $\{c=1 \rightarrow x, c=2\rightarrow \lambda y . x, c=3 \rightarrow (x\;y)\}$. This symbolic union is depicted in Figure~\ref{fig:syntax-tree-repr}. Larger ASTs can then be created recursively using the same process, using smaller ASTs as subtrees.

Notice that the structure of these trees depends on the syntax of the language. We thus call such trees ``syntax trees." Syntax trees are appealing because of their simplicity: indeed, they are the most commonly-used symbolic representations of ASTs. However, they can become extremely inefficient when symbolically evaluated.

\paragraph{Scalability challenges}

\begin{figure}
\begin{tabular}{p{0.5\columnwidth}|p{0.5\columnwidth}}
\includegraphics[width=0.5\columnwidth]{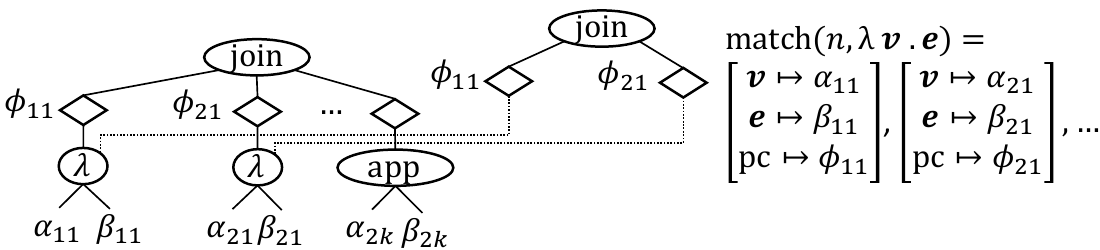}
&
\includegraphics[width=0.5\columnwidth]{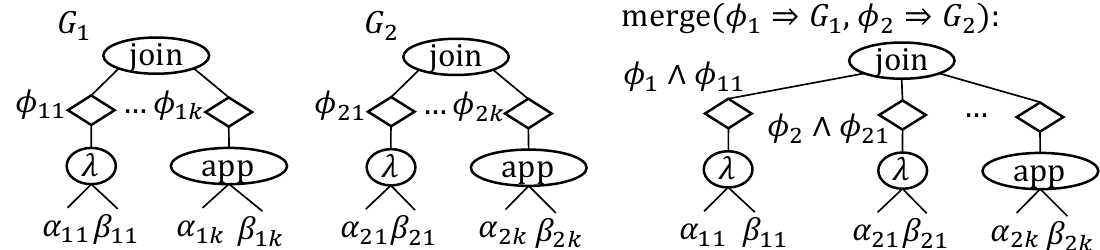}
\end{tabular}
%\nocaptionrule
\caption{Left: A ``match" operation on a symbolic syntax tree. Right: A ``merge" operation on two symbolic syntax trees.}
\label{fig:syntax-tree-bnm}
\end{figure}

The inefficiency of symbolic syntax trees can be revealed by analyzing their ``branch-and-merge" operations. Recall that the ``branch" operation takes place on a symbolic condition: in the case of an operation on ASTs, this condition determines whether a tree matches a given pattern. To perform a pattern-match on a symbolic syntax tree, the symbolic evaluator must examine the members of the symbolic union and individually check whether \emph{each} member matches. Furthermore, when pattern-matching, we usually wish to extract subtrees that match our pattern's metavariables. The symbolic evaluator must therefore perform this extraction on each member of the symbolic union and then merge the results into a \emph{fresh} symbolic union. This situation is depicted in Figure~\ref{fig:syntax-tree-bnm} on the left. For a symbolic union with $n$ members, this operation takes at least $O(n)$ time.

Analyzing the ``merge" operation reveals that such a merge often grows the size of the symbolic union. To merge two symbolic unions, a symbolic evaluator must add together the members of each symbolic union, updating the respective path conditions as required. In the worst case, this can double the size of the symbolic union. This is depicted in Figure~\ref{fig:syntax-tree-bnm} on the right.
Over the course of many branch-and-merge operations, the symbolic unions that represent symbolic syntax trees grow. Over time, symbolic evaluation slows down, and ultimately becomes impractically inefficient.

\paragraph{Possible solutions}

A crucial insight is that syntax trees are inefficient because their structure depends on the grammar of the language. Representing a set of differently-shaped trees requires us to create a symbolic union, which is expensive to match against. This intuition guides us to the key idea behind the Bonsai tree: maintain a single tree that may or may not be syntactically-valid, and push the constraints of the syntax to assertions in the assertion store. Such a ``clean" data structure, unencumbered by syntactic constraints, should be efficiently manipulable by a symbolic evaluator. In the remainder of this paper, we demonstrate that this is indeed the case.

\section{Bonsai Trees}
\label{sec:bonsai-trees}

\begin{figure}[t]
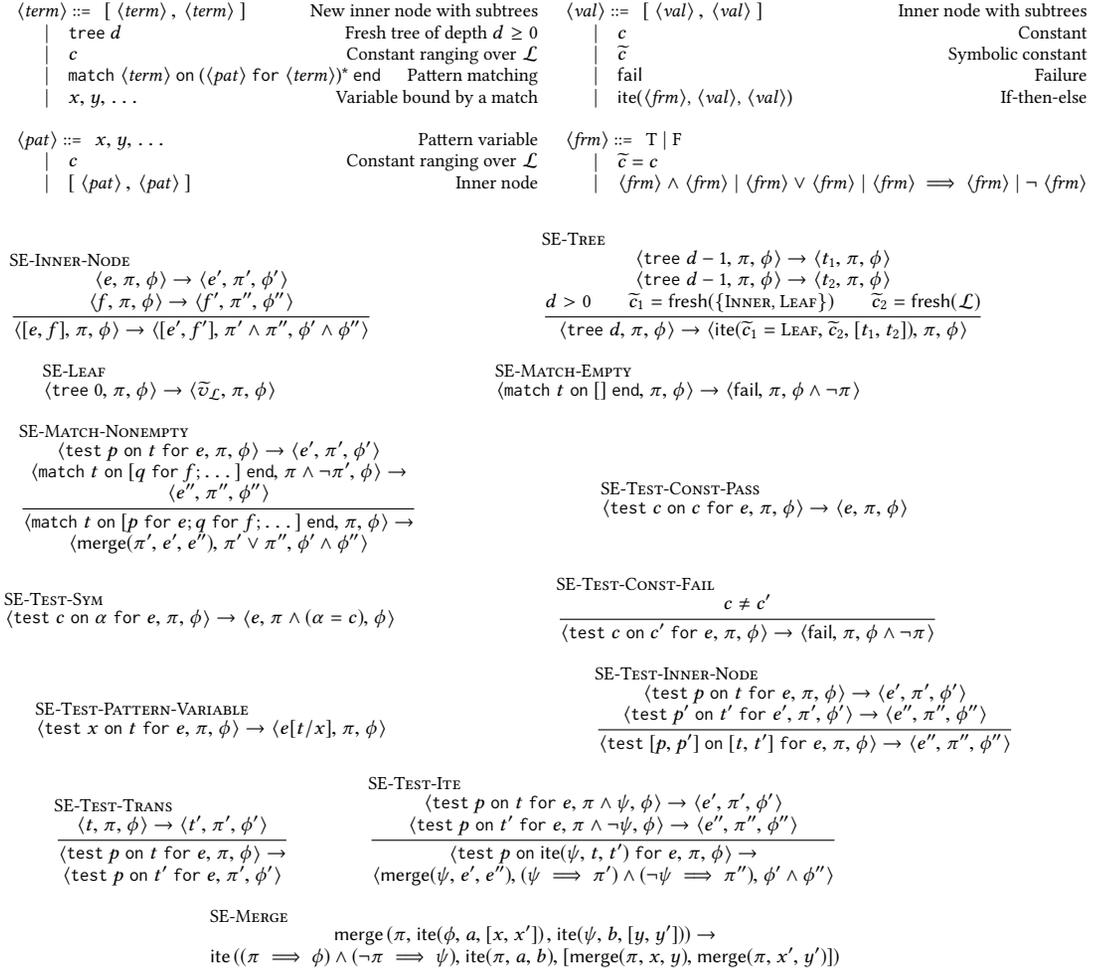

\scriptsize
% 12:21am black magic
\renewcommand{\DefTirName}[1]{\textsc{SE-#1}}
\newcommand{\pr}[1]{\left<{#1}\right>}
\begin{tabular}{p{0.5\columnwidth}p{0.5\columnwidth}}
\begin{grammar}
<term> ::=
$[$ <term> $,$ <term> $]$ \hfill New inner node with subtrees
\alt "tree" $d$ \hfill Fresh tree of depth $d \geq 0$
\alt $c$ \hfill Constant ranging over $\mathcal{L}$
\alt "match" <term> "on" (<pat> "for" <term>)* "end" \hfill Pattern matching
\alt $x, y, \dots$ \hfill Variable bound by a match

<pat> ::= $x, y, \dots$ \hfill Pattern variable
\alt $c$ \hfill Constant ranging over $\mathcal{L}$
\alt $[$ <pat> $,$ <pat> $]$ \hfill Inner node
\end{grammar}
&
\begin{grammar}
<val> ::=
$[$ <val> $,$ <val> $]$ \hfill Inner node with subtrees
\alt $c$ \hfill Constant
\alt $\widetilde{c}$ \hfill Symbolic constant
\alt $\mathsf{fail}$ \hfill Failure
\alt $\textsf{ite}$(<frm>, <val>, <val>) \hfill If-then-else

<frm> ::= T | F
\alt $\widetilde{c}$ $=$ $c$
\alt <frm> $\wedge$ <frm> | <frm> $\vee$ <frm> | <frm> $\implies$ <frm> | $\neg$ <frm>
\end{grammar}
\end{tabular}

\begin{mathpar}
\inferrule[Inner-Node]
{
\pr{e, \pi, \phi}
\rightarrow
\pr{e^\prime, \pi^\prime, \phi^\prime}
\\\\
\pr{f, \pi, \phi}
\rightarrow
\pr{f^\prime, \pi^{\prime\prime}, \phi^{\prime\prime}}
}
{
\pr{[e, f], \pi, \phi}
\rightarrow
\pr{[e^\prime, f^\prime],
\pi^\prime \wedge \pi^{\prime\prime},
\phi^\prime \wedge \phi^{\prime\prime}}
}

\and

\inferrule[Tree]
{
\pr{\lit*{tree}\;d-1, \pi, \phi}
\rightarrow
\pr{t_1, \pi, \phi}
\\\\
\pr{\lit*{tree}\;d-1, \pi, \phi}
\rightarrow
\pr{t_2, \pi, \phi}
\\\\
d>0
\\
\widetilde{c}_1 = \mathsf{fresh(\textsc{\{Inner, Leaf\}})} \\
\widetilde{c}_2 = \mathsf{fresh(\mathcal{L})}
}{
\pr{\lit*{tree}\;d, \pi, \phi}
\rightarrow
\pr{\mathsf{ite}(\widetilde{c}_1 = \mathsf{\textsc{Leaf}},
\widetilde{c}_2,
[t_1, t_2]
), \pi, \phi}
}

\and

\inferrule[Leaf]
{
\pr{\lit*{tree}\;0, \pi, \phi}
\rightarrow
\pr{\widetilde{v}_\mathcal{L}, \pi, \phi}
}{}

\and

\inferrule[Match-Empty]
{
\pr{\lit*{match}\;t\;\lit*{on}\;[]\;\lit*{end}, \pi, \phi}
\rightarrow
\pr{\mathsf{fail}, \pi, \phi\wedge\neg\pi}
}{}

\and

\inferrule[Match-Nonempty]
{
\pr{\texttt{test}\;p\;\texttt{on}\;t\;\texttt{for}\; e, \pi, \phi}
\rightarrow
\pr{e^\prime, \pi^\prime, \phi^\prime}
\\\\
\pr{\lit*{match}\;t\;\lit*{on}\; [q\; \texttt{for}\; f; \dots] \;\lit*{end},
\pi\wedge\neg\pi^\prime, \phi}
\rightarrow \\\\
\pr{e^{\prime\prime}, \pi^{\prime\prime}, \phi^{\prime\prime}}
}{
\pr{\lit*{match}\;t\;\lit*{on}\;
[p \;\texttt{for}\; e;
 q \;\texttt{for}\; f; \dots]\;\lit*{end}, \pi, \phi}
\rightarrow\\\\
\pr{\mathsf{merge}(\pi^\prime, e^\prime, e^{\prime\prime}),
\pi^\prime \vee \pi^{\prime\prime},
\phi^\prime \wedge \phi^{\prime\prime}
}
}

\and

\inferrule[Test-Const-Pass]
{
\pr{\texttt{test}\;c\; \texttt{on}\;c\; \texttt{for}\; e, \pi, \phi}
\rightarrow
\pr{e, \pi, \phi}
}{}

\and

\inferrule[Test-Sym]
{
\pr{\texttt{test}\;c\; \texttt{on}\;\alpha\; \texttt{for}\; e, \pi, \phi}
\rightarrow
\pr{e, \pi \wedge (\alpha = c), \phi}
}{}

\and

\inferrule[Test-Const-Fail]{
c \not = c^\prime
}{
\pr{\texttt{test}\;c\; \texttt{on}\;c^\prime\; \texttt{for}\; e, \pi, \phi}
\rightarrow
\pr{\mathsf{fail}, \pi, \phi \wedge \neg \pi}
}

\and

\inferrule[Test-Pattern-Variable]
{
\pr{\texttt{test}\; x \; \texttt{on}\;t\; \texttt{for}\; e, \pi, \phi}
\rightarrow
\pr{e[t/x], \pi, \phi}
}{}

\and

\inferrule[Test-Inner-Node]
{
\pr{\texttt{test}\;p\;\texttt{on}\;t\;\texttt{for}\; e, \pi, \phi}
\rightarrow
\pr{e^\prime, \pi^\prime, \phi^\prime}
\\\\
\pr{\texttt{test}\;p^\prime\; \texttt{on}\;t^\prime\; \texttt{for}\;e^\prime,\pi^\prime, \phi^\prime}
\rightarrow
\pr{e^{\prime\prime}, \pi^{\prime\prime}, \phi^{\prime\prime}}
}{
\pr{\texttt{test}\;[p, p^\prime]\; \texttt{on}\;[t, t^\prime]\; \texttt{for}\; e, \pi, \phi}
\rightarrow
\pr{e^{\prime\prime}, \pi^{\prime\prime}, \phi^{\prime\prime}}
}

\inferrule[Test-Trans]
{
\pr{t, \pi, \phi}
\rightarrow
\pr{t^\prime, \pi^\prime, \phi^\prime}
}{
\pr{\texttt{test}\;p\;\texttt{on}\;t\;\texttt{for}\; e, \pi, \phi}
\rightarrow\\\\
\pr{\texttt{test}\;p\;\texttt{on}\;t^\prime\;\texttt{for}\; e, \pi^\prime,\phi^\prime}
}

\inferrule[Test-Ite]
{
\pr{\texttt{test}\;p\;\texttt{on}\;t\;\texttt{for}\; e, \pi \wedge \psi, \phi}
\rightarrow
\pr{e^\prime, \pi^\prime, \phi^\prime}
\\\\
\pr{\texttt{test}\;p\;\texttt{on}\;t^\prime\;\texttt{for}\; e, \pi \wedge \neg\psi, \phi}
\rightarrow
\pr{e^{\prime\prime}, \pi^{\prime\prime}, \phi^{\prime\prime}}
}{
\pr{\texttt{test}\; p \;\texttt{on}\; \mathsf{ite}(\psi, t, t^\prime) \;\texttt{for}\; e, \pi, \phi}
\rightarrow \\\\
\pr{\mathsf{merge}(\psi, e^\prime, e^{\prime\prime}),
(\psi \implies \pi^\prime ) \wedge (\neg\psi \implies \pi^{\prime\prime}),
\phi^\prime \wedge \phi^{\prime\prime}}
}

\and

\inferrule[Merge]{
\mathsf{merge}\left( \pi, \mathsf{ite}(\phi, a, [x, x^\prime]\right),
\mathsf{ite}(\psi, b, [y, y^\prime]))
\rightarrow\\\\
\mathsf{ite} \left( (\pi \implies
\phi) \wedge (\neg\pi \implies \psi), \mathsf{ite}(\pi, a, b),
[\mathsf{merge}(\pi, x, y), \mathsf{merge}(\pi, x^\prime, y^\prime)] \right)
}{}

\end{mathpar}

\caption{The \toolname\ Core language and reduction rules.
Here,
$t, e, f$ range over terms;
$p, q$ range over patterns;
$x, y$ range over pattern variables;
$c$ ranges over $\mathcal{L}$, the set of possible leaf values;
$\alpha$ ranges over $\mathcal{S}$, the set of symbolic constants;
$\pi, \phi, \psi$ range over boolean formulas with equality defined on $\mathcal{S}\times\mathcal{L}$.
$\pr{t, \pi, \phi}$ denotes term $t$ with path condition $\pi$ and assertion store $\phi$.
\textsf{fresh($\mathcal{A}$)} denotes a fresh symbolic constant ranging over set $\mathcal{A}$.
$\mathsf{merge}$ is defined in its most general form: a merge of two \textsf{ite} values. Definitions for merging non-\textsf{ite} values can be derived by setting $\phi$ and/or $\psi$ to \textsf{T}/\textsf{F}.
}
\label{fig:bonsai-core}
\end{figure}

We now formalize the intuition above in terms of the \emph{Bonsai Tree}, an efficient alternative to the syntax tree. Our discussion is guided by the \toolname\ Core language, a minimal calculus that describes programs that manipulate symbolic \toolname\ trees via ``match" and ``merge" operations (Figure~\ref{fig:bonsai-core}). An empirical evaluation of the Bonsai encoding is presented in Section~\ref{sec:eval:scalability}.

\paragraph{Structure}

\begin{figure}
\begin{tabular}{p{0.5\columnwidth}|p{0.5\columnwidth}}
\includegraphics[width=0.5\columnwidth]{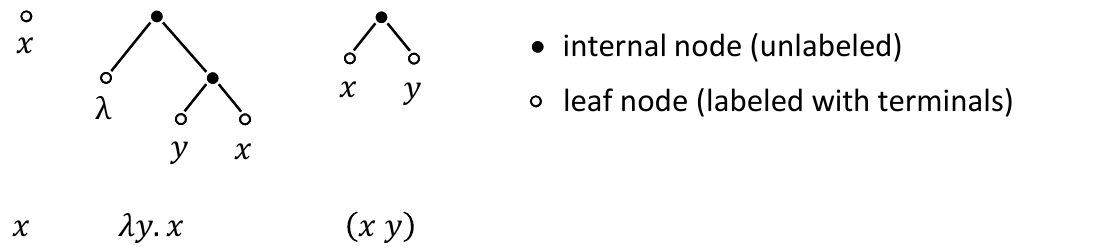}
&
\includegraphics[width=0.5\columnwidth]{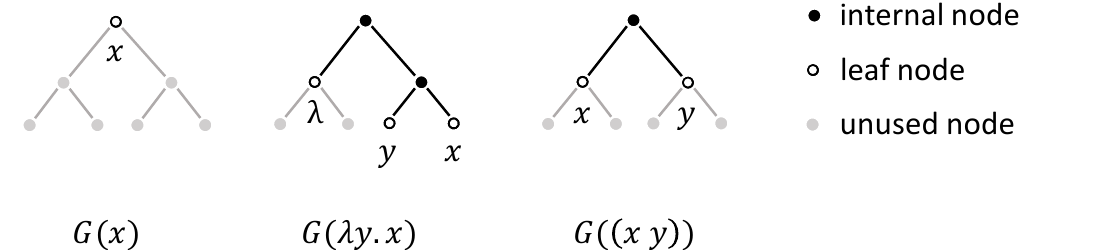}
\end{tabular}
%\nocaptionrule
\caption{Left: Concrete Bonsai trees for terms $x$, $\lambda y.x$, and $(x \; y)$. Right: and their embeddings in perfect binary trees.}
\label{fig:bonsai-tree-repr}
\end{figure}

\begin{figure}
\centering\includegraphics[width=0.5\columnwidth]{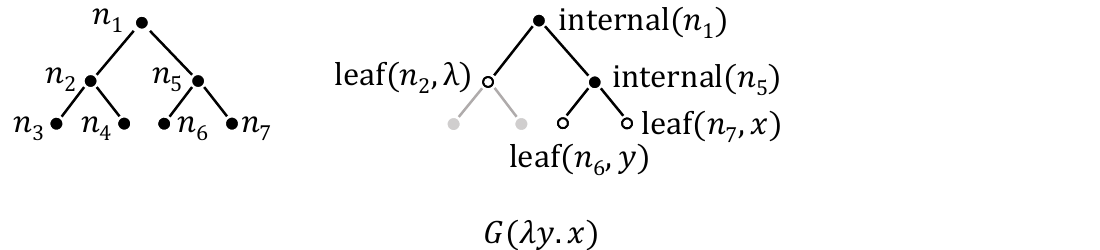}
%\nocaptionrule
\caption{The representation of the embedded tree for $\lambda y.x$.}
\label{fig:bonsai-tree-pred}
\end{figure}

Bonsai trees are embedded in perfect binary trees. Figure~\ref{fig:bonsai-tree-repr} depicts three such ``concrete" trees, and their embeddings in perfect binary trees. As shown in the \toolname\ Core rule \textsc{SE-Tree}, the embedding is represented by assigning two symbolic constants to each node of the binary tree. The first determines whether the node is internal or a leaf, and the second determines the terminal in case it is a leaf. Figure~\ref{fig:bonsai-tree-pred} depicts one such representation.

More formally, a \emph{Bonsai tree} $G$ is a tuple $\left< T, \mathcal{L}, m, \textsf{internal}, \textsf{leaf} \right>$ where $T=\left<N,E\right>$ is a perfect binary tree of  depth $d$ and size $m=2^d-1$, such that $N$ is the set of nodes $\{n_1,\ldots,n_{m}\}$ and $\mathcal{L}$ is the set of potential leaf symbols. The predicates $\textsf{internal}: N\rightarrow \text{Bool}$ and  $\textsf{leaf}: (N \times \mathcal{L}) \rightarrow \text{Bool}$ specify what tree is embedded in $G$. In a \emph{symbolic} Bonsai tree, these guards are logical formulas composed of symbolic constants.

\paragraph{Branch-and-merge operations}

Bonsai carefully maintains the following invariant: each \textsf{ite} created in the language has the property that if its condition is true, the result is a leaf, else it is an inner node. This invariant can easily be verified for the rules \textsc{SE-Tree} and \textsc{SE-Merge}, the only sources of \textsf{ites} in Bonsai Core.

\begin{figure}
\begin{tabular}{p{0.5\columnwidth}|p{0.5\columnwidth}}
\includegraphics[width=0.5\columnwidth]{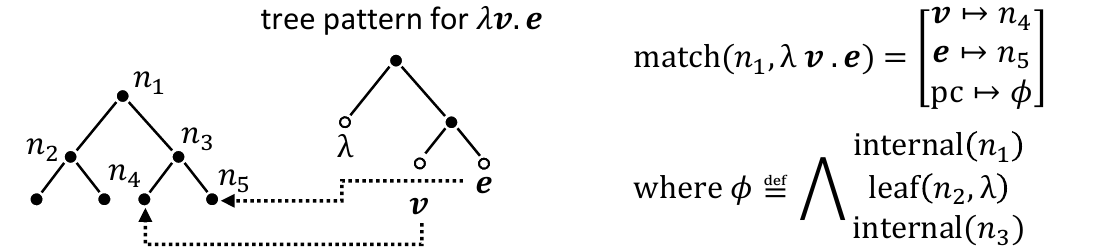}
&
\includegraphics[width=0.5\columnwidth]{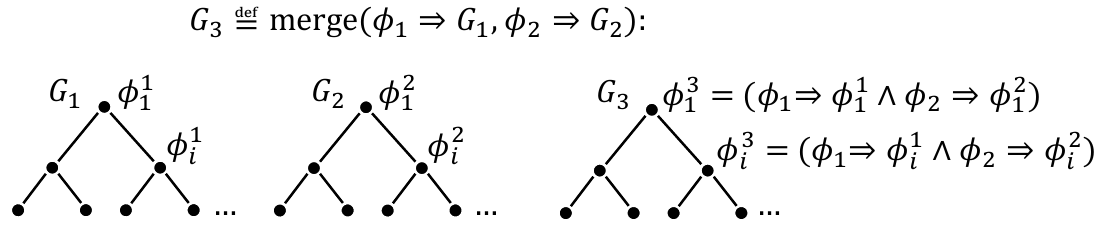}
\end{tabular}
%\nocaptionrule
\caption{Left: Matching on a Bonsai tree guarantees to return a single match, at the cost of creating a larger formula. Right: Merging two Bonsai trees. At the constant cost of merging a guard for each node, we avoid growing the tree.}
\label{fig:bonsai-tree-bnm}
\end{figure}

A useful result of this invariant is that a pattern matcher need not branch when matching against an \textsf{ite}: depending on whether the pattern is a leaf or an inner node, it is clear which path of the \textsf{ite} must be taken. Thus, symbolic Bonsai trees can be matched in a single operation. In contrast, recall that symbolic syntax trees required a match operation for \emph{each} member of their symbolic unions. Pattern-matching a Bonsai tree is depicted in Figure~\ref{fig:bonsai-tree-bnm} on the left. The path condition $\phi$ represents the conditions required for that Bonsai tree to match the pattern. Notice that since the pattern-matcher never branches, any subtree can be extracted directly. In contrast, the symbolic syntax tree must merge the extracted subtrees from each individual match into another, larger symbolic union.

A merge of two Bonsai trees under path conditions $\phi_1, \phi_2 := \neg\phi_1$ is depicted in Figure~\ref{fig:bonsai-tree-bnm} on the right. No matter what the guards for these individual trees are, we can merge them by joining them node-wise under these path conditions. Thus, though we require $O(m)$ operations to merge trees of size $m$, the trees never grow: the output of the merge is the same size as each input tree, and the $m$ operations are a constant time cost. In contrast, the symbolic syntax tree grow at each merge, and thus the time cost grows as more branch-and-merge operations are carried out.

Notice that while the tree itself does not grow, the \emph{formulas} for the guards do indeed grow at each merge due to the addition of path conditions. This, however, is not a concern for two reasons.
First, symbolic evaluation engines represent formulas as DAGs. Creating a disjunction of two formulas does not require us to duplicate each formula in memory, which would be an expensive operation. Instead, we only need to allocate a fresh ``$\vee$" operator node, and create cheap pointers to existing formulas for the operands.
Second, we never have to ``look inside" these formulas. We only construct them and, when finished, pass them on to the SMT solver. As a result, the growing formulae do not affect the efficiency of the symbolic evaluation.

\paragraph{Example}

\begin{figure}
    \centering
    \includegraphics[width=0.9\columnwidth]{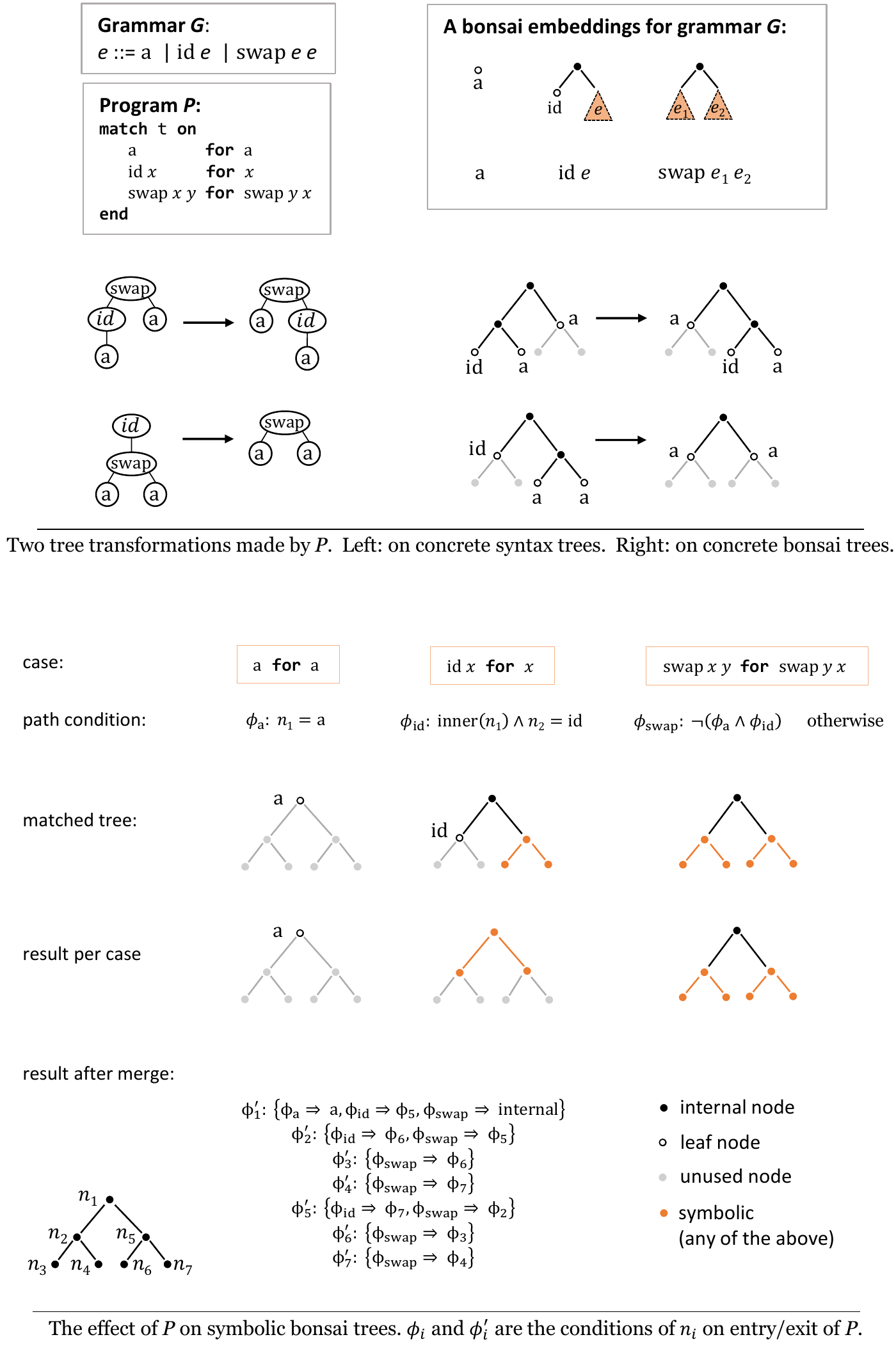}
    \caption{An example of pattern-matching against a symbolic \toolname\ tree and merging the results.}
    \label{fig:arithmetic-example}
\end{figure}

Figure~\ref{fig:arithmetic-example} depicts a small worked example of pattern-matching against and merging symbolic \toolname\ trees. The mini-language has three terms: \texttt{id(x)}, which evaluates to its argument, \texttt{swap(x, y)} which swaps its arguments when evaluated, and \texttt{a}, a constant that evaluates to itself. The grammar, as well as its embedding in binary trees, is presented in the top half of the figure. The bottom half depicts the symbolic evaluation of the \toolname\ Core term
\begin{quote}
\footnotesize
\begin{verbatim}
match (tree 3) on
    a for a
    [swap, [x, y]] for [swap, [y, x]]
    [id, x] for x
end
\end{verbatim}
\end{quote}
We first perform each individual match, creating path conditions and results, and then merge the results under each path condition.
\section{Implementation}
\label{sec:implementation}

This section illustrates how to implement \toolname\ trees in a specific language, Racket~\cite{plt-tr1}, using a specific symbolic evaluator, Rosette~\cite{rosetteguide}. Of course, guided by the rules in Figure~\ref{fig:bonsai-core}, these ideas can be implemented in other symbolic evaluation platforms as well~\cite{DBLP:conf/sigsoft/SenKBG13,DBLP:conf/sigsoft/SenNGC15}. After implementing \toolname\ trees, we demonstrate how to use them to check the arithmetic language in Figure~\ref{fig:argrammar}.

\paragraph{Constructing a symbolic tree}
Recall rule \textsc{SE-Tree} from Figure~\ref{fig:bonsai-core}.  Two fresh symbolic constants on each node dictate whether it is an internal node or a leaf, and if it is a leaf, which symbol it represents. We can construct such a node by symbolically evaluating the call to  \texttt{\small binary-tree!} shown in Figure~\ref{fig:binaryencoding} (left). The call is recursively unrolled by the symbolic evaluator until the depth limit reaches zero, at which point we follow \textsc{SE-Leaf} by creating a single symbolic constant to represent which symbol is represented by the leaf. \texttt{fresh-symbolic!} is a macro that symbolically selects from a set $S$, analogous to $\mathsf{fresh}(S)$ in \toolname\ Core. Internally, it is implemented by creating symbolic integer constants and maintaining a map from integers to members of $S$. Inner nodes are represented by \texttt{cons} pairs so that we can take advantage of existing Racket primitives for manipulating S-expressions.

The result of symbolically evaluating \texttt{\small binary-tree!} is a data structure that embeds the symbolic constants $s_i$, pictured in Figure~\ref{fig:binaryencoding} (right). Note that this tree contains no syntactic information: concrete assignments of values to each $s_i$ may create syntactically-invalid ASTs.

\begin{figure}[t]
\begin{tabular}{p{7cm}p{7cm}}
\footnotesize
\begin{verbatim}
(define (binary-tree! depth)
  (if (> depth 0)
      ; SE-Tree
      (if
       (eq? 'leaf (fresh-symbolic! '(leaf inner)))
       (fresh-symbolic! all-leaves)
       (cons
          (binary-tree! (- depth 1))
          (binary-tree! (- depth 1))))
      ; SE-Leaf
      (fresh-symbolic! all-leaves)))

(define all-leaves '(zero succ zero? if))
; for arithmetic language
\end{verbatim}
&
\footnotesize
\begin{verbatim}
(ite (= s1 leaf)
     s2
     (cons (ite (= s2 leaf) s4 ...)
           (ite (= s3 leaf) s5 ...)))
\end{verbatim}
\end{tabular}
\caption{A procedure that generates a fresh symbolic Bonsai tree (left), and its output (right).}
\label{fig:binaryencoding}
\end{figure}

\paragraph{Pattern-matching}
The \toolname\ pattern-matching macro, \texttt{tree-match}, is implemented by following the rules
\textsc{SE-Test-\{
Const-Pass, Const-Fail, Sym, Pattern-Variable, Inner-Node, Trans
\}}
as expected. Since the pattern-matcher is written within a symbolically-evaluated language, \textsc{SE-Test-Ite} follows automatically from the symbolic evaluator's behavior on operations on \textsf{ites}. Figure~\ref{fig:syntax-constraints} illustrates the behavior of the pattern-matcher using the example of syntactic constraints (see below).
Note that as per \textsc{SE-Match-Nonempty}, the pattern-matcher is sensitive to the order in which patterns are presented. If a tree matches two patterns $p_1$ and $p_2$, the earlier one ($p_1$) is given priority. The path condition for the $p_2$ match will stipulate that the $p_1$ match fails. This resolves ambiguities in case the patterns are not mutually exclusive.

\paragraph{Syntactic constraints}
To assert that the \toolname\ tree embeds only syntactically correct trees, we can take advantage of pattern-matching. We invoke \texttt{\small (assert (is-program? tree))}, which traverses the tree and generates constraints required by the grammar defined in Figure~\ref{fig:argrammar}. This call expands into the procedures in Figure~\ref{fig:syntax-constraints}. More generally, we provide the macro \texttt{syntax-matches?}, which converts a grammar in BNF to a syntax checker in the style of \texttt{is-program?}.

\begin{figure}
\begin{tabular}{p{6cm}p{6cm}}
\footnotesize
\begin{verbatim}
(define (is-program? p)
  (tree-match p
    'zero (lambda () #t)
     
    '(succ _) (lambda (x) (is-program? x))
      
    '(if _ _ _) (lambda (c t f)
                  (and (is-program? c)
                      (is-program? t)
                      (is-program? f)))
           
    '(zero? _) (lambda (x) (is-program? x))
      
    '_ (lambda () #f)))
\end{verbatim}
&
\footnotesize
\begin{verbatim}
(define (is-program? p)
  (cond [(eq? p 'zero) #t]
        [(and (pair? p)
              (eq? (car p) 'succ)
              (null? (cddr p)))
         (is-program? (cadr p))]
        [(and (pair? p)
              (eq? (car p) 'if)
              (pair? (cdr p))
              (pair? (cddr p))
              (pair? (cdddr p))
              (null? (cddddr p)))
         (and (is-program? (cadr p))
              (is-program? (caddr p))
              (is-program? (cadddr p)))]
        [(and (pair? p)
              (eq? (car p) 'zero?)
              (null? (cddr p)))
         (is-program? (cadr p))]))
\end{verbatim}
\end{tabular}
\caption{Placing syntactic constraints on the Bonsai tree.  The code results from macro expanding the arithmetic grammar (left).  One more level of expansion (right) reveals the internal Bonsai tree structure.}
\label{fig:syntax-constraints}
\end{figure}

\paragraph{Merge}
Recall from Section~\ref{sec:symex} that symbolic evaluators already perform merges on conditional branches. Rosette in particular also performs some simplification: when merging two \texttt{cons} pairs, it first individually merges the heads and tails, and then creates a fresh \texttt{cons} pair of the two merges. This gives us half of the merge operation ``for free": we must only be careful to arrange for Rosette to merge pairs separately from leaves, since the simplification does not apply to merging a leaf with an inner node. This, however, is always the case as long as the ites being merged satisfy the invariant described in Section~\ref{sec:bonsai-trees}. Thus, by carefully regulating the creation of \texttt{ite} values in \texttt{binary-tree!}, we force Rosette to always perform a Bonsai merge on trees.

\subsection{The \toolname\ Checker}

Equipped with the \toolname\ Tree, we now return to our goal of symbolically reasoning about the language in Figure~\ref{fig:argrammar}. We begin by searching for counterexamples to soundness. The \toolname\ Tree allows us to efficiently execute the algorithm shown earlier in Figure~\ref{fig:venn-diagram}. First, create a symbolic representation of the set of all ASTs up to some maximum size $m$. Next, compute symbolic representations of trees that (a)~are syntactically valid; (b)~pass the typechecker; (c)~fail in the interpreter. Finally, ask the solver to find a tree in the intersection of these sets.  If it exists, this program is a counterexample.

Recall that the typechecker in Figure~\ref{fig:argrammar} has a soundness bug, failing to verify that the two clauses of `if' expressions have the same type. \toolname\ catches this error by executing the algorithm described above:
\begin{quote}
\footnotesize
\begin{verbatim}
(define program (binary-tree! 10))     
(assert-pass (syntax-matches? arithmetic-syntax program))
(assert-pass (check program))                   
(assert-fail (execute program))                 
(define solution-model (solve))                 
(echo (evaluate program solution-model))        
\end{verbatim}
\end{quote}
The first line creates the formula characterizing the set $A$ of all abstract syntax trees of depth 10 or less that adhere to the grammar. The second line creates the formula $S$, which describes the subset of programs in $A$ that are syntactically-valid. The third line evaluates the formula on the typechecker to create the formula $T$, which describes the set of programs that pass the typechecker. The fourth line does the same with the interpreter, negating the formula generated to produce $I^\prime$. The fifth line queries the solver for a model that satisfies the formula $S \wedge T \wedge I^\prime$, and the sixth line simply interprets the model to yield a concrete program, which is then echoed to the user.
\toolname\ returns the following counterexample in less than 2~seconds.
\begin{quote}
\footnotesize
\begin{verbatim}
'(succ (if (zero? (succ zero)) zero (zero? zero)))
\end{verbatim}
\end{quote}
Indeed, this program passes the typechecker but causes the interpreter to crash by attempting to increment a boolean value. If we reintroduce the omitted check in our typechecker, our system fails to find a counterexample, showing that this check fixes the bug.

\subsection{A Variation on the Query}

A small variation on the counterexample-finding algorithm above allows us to compute a ``diff" of two typecheckers by querying for programs that are accepted by one but rejected by the other. Suppose a student submits the unsound arithmetic typechecker above to a teacher. The teacher can compare it to a correct reference implementation and give the student feedback in the form of a program accepted by the former but rejected by the latter.

To synthesize such a program, we initialize $T$, the set of all programs that pass the student's implementation, and $U$, the set of all programs that pass the teacher's implementation. Then, we query the SAT solver for any member of the set $S \wedge T \wedge U^\prime$.
\begin{quote}
\footnotesize
\begin{verbatim}
(define program (binary-tree! 10))
(assert-pass (syntax-matches? arithmetic-syntax program))
(assert-pass (student-check program))
(assert-fail (teacher-check program))
(define solution-model (solve))
(echo (evaluate program solution-model)) 
; ==> '(if (iszero? zero) zero (iszero? zero))
\end{verbatim}
\end{quote}
Note that this program is not a counterexample to \emph{soundness} because it does not fail at runtime. However, it behaves differently on the two typecheckers, and thus represents a member of their ``diff."

\section{Case Studies}
\label{sec:casestudies}

We now demonstrate how \toolname\ can be used to explore three more advanced type systems: an object-oriented, an ownership-based, and a dependent type system. Besides looking for soundness issues, we also run queries that help us better understand the type system's restrictions.

%
% ,__                 .    _                                           _       .   
% /  `   ___    ___  _/_   /        ___  .___  ,  _  /   ___  `   ___. /      _/_  
% |__  .'   `  /   `  |    |,---. .'   ` /   \ |  |  | .'   ` | .'   ` |,---.  |   
% |    |----' |    |  |    |'   ` |----' |   ' `  ^  ' |----' | |    | |'   `  |   
% |    `.___, `.__/|  \__/ /    | `.___, /      \/ \/  `.___, /  `---| /    |  \__/
% /                                                              \___/             
%

\subsection{Featherweight Java}

Featherweight Java~\cite{featherweightjava} is a calculus that models essential features of the Java programming language. Featherweight Java has classes with methods, as well as inheritance via subclasses.

We test \toolname's ability to identify the classical bug in function subtyping, which is to insist that function arguments are covariant, i.e. to assume that $s <: t$ implies $(s \rightarrow u) <: (t \rightarrow u)$. The correct approach is to require contravariance in function arguments, i.e. if $s <: t$ then $(s \rightarrow u) :> (t \rightarrow u)$. This bug is of historical interest because it appeared in early versions of the Eiffel programming language, and its discovery came as a surprise to the Eiffel community~\cite{eiffelbug}.

After we introduced the bug into our Featherweight Java implementation, \toolname\ reported the following counterexample in a few seconds. This counterexample has been manually formatted to Java syntax:
\begin{quote}
\footnotesize
\begin{verbatim}
class A extends Object {
    B bar(A arg) {
        return arg.bar(this);
    } }
class B extends A {
    B quux(C arg) {
        return new A().bar(arg);
    } }
class C extends B {
    B bar(B arg) {
        return arg.quux(this);
    } }
new C().quux(new C());
\end{verbatim}
\end{quote}
The call stack execution trace:
\begin{quote}
\footnotesize
\begin{lstlisting}
Error: class A has no method quux
in new$_2$ A().quux(new$_1$ C()); // line 11
...new$_1$ C().bar(new$_2$ A());  // line  3
...new$_2$ A().bar(new$_1$ C());  // line  7
...new$_0$ C().quux(new$_1$ C()); // line 13
...[init]
\end{lstlisting}
\end{quote}
In the stack trace, we can see that the failure is caused by \texttt{new C().bar(new A())}. 
This call occurs because the method \texttt{C.bar} is unsoundly allowed to be used in place of \texttt{A.bar} because of their covariant argument types.
This explains why the covariant argument rule is problematic: whenever \texttt{C <: A}, the rule promises that we can safely use a \texttt{C} wherever we use an \texttt{A}. This promise is violated because \texttt{A.bar} accepts inputs that \texttt{C.bar} does not.
Note that this issue does not apply to true Java, because shadowing an inherited method with a different argument type overloads that method name, creating a distinct method.

The symbolic evaluation of Featherweight Java is much more demanding than that of the arithmetic language. To typecheck and evaluate Featherweight Java, we must make lookups into a class table in order to traverse the subclass hierarchy; when the class table is symbolic, a lookup could potentially be any of the entries in the table. Representing the class table as a \toolname\ Tree allows us to efficiently merge all possible results of a lookup into small, compact formulas that can easily be manipulated by the interpreter.

%                                          
%                               |    o     
% ,---.. . .,---.,---.,---.,---.|---..,---.
% |   || | ||   ||---'|    `---.|   |||   |
% `---'`-'-'`   '`---'`    `---'`   '`|---'
%                                     |    
% 

\subsection{Ownership Java}
\label{sec:ownershipjava}

Our next language augments Featherweight Java with an object-oriented ownership type system, Ownership Java~\cite{ownershipjava, safejava}.
The safety guarantee made by Ownership Java is that well-typed objects must access only objects that they \emph{own}. The type system statically enforces this encapsulation guarantee by keeping track of the binary ownership relation on objects, which is declared in annotations on each class, method, and field. 

A canonical example of Ownership Java is the implementation of a stack. The stack object is allowed to access its inner linked list but not the objects stored in the linked list; those can only be accessed by the client of the stack.  The stack class is thus defined with two owner parameters. The first parameter is special because it is the owner of the stack object at runtime. The second owner parameter denotes the client who owns the data in the list; this parameter is used by the stack class implementation to annotate its methods and the encapsulated linked list.  For example, the return value from the \texttt{pop} method is annotated with the second owner parameter, allowing the client to access objects retrieved from the stack.

To prove soundness of Ownership Java, \citet{ownershipjava} introduce an ``ownership tree," a transitive view of the ownership relation: $o_2$ is a direct descendant of $o_1$ if $o_1$ owns $o_2$. The ownership tree is rooted at the special owner \texttt{world}.  Using the ownership tree, the \emph{encapsulation theorem} is stated: object $x$ can access object $o$ only if (1)~$x = o$, (2)~$x$ is a descendant of $o$ in the ownership tree, or (3)~$x$ is an inner class object of $o$. This invariant can be checked by tools like Pipal (Section~\ref{sec:relatedwork}).

With the \toolname\ Checker, however, we do not need to formulate any such invariant. Instead, we only need one straightforward dynamic check to detect unsoundness. The check asserts that the owner stored in the accessed object (\texttt{object}) is owned by the receiver of the accessor method (\texttt{this}):
\begin{quote}
\footnotesize
\begin{verbatim}
(define (evaluate-expression ...)
    ...
    (assert
        (owns? this object)
        "Error: attempted invalid access!")
    ...)
\end{verbatim}
\end{quote}
This check is easy to implement because it is a direct statement of the desired guarantee.

We believe that a counterexample that violates this assertion provides a more intuitive understanding of the issue than a counterexample that violates the encapsulation theorem because a trace shows the origins of the unsafe state. We reproduce one such counterexample below, found by \toolname\ in about 90 seconds when we disabled the static owner check of method calls:
\begin{quote}
\footnotesize
\begin{verbatim}
class Main<O1, O2> extends Foo {
    Main<O1, O1> meth3(Main<O1, O1> arg) {
        (arg.meth3(arg)).main(arg);
    } }
class Foo<O1, O2> extends Object {
    Main<O2, O2> main(Main<O2, O2> arg) {
        (new Main<O2, O2>()).meth3(arg);
    } }
new Main<world, world>().main(new Main<world, world>());
\end{verbatim}
\end{quote}
This program fails at runtime because \texttt{Foo<O1, O2>} should not be allowed to access methods of a \texttt{Main<O2, O2>} on line 7. It should only be allowed to access an instance of \texttt{Main} that it owns, which could be a \texttt{Main<this, O1>} or a \texttt{Main<this, world>}.

\paragraph{Why have additional constraints on owners?}

Next, we illustrate \toolname's ability to compare type systems by formulating queries involving multiple typecheckers and interpreters. 

When instantiating a new object with owner parameters $o_{0\dots n}$, the Ownership Java type system insists that the object's owner $o_0$ is a descendant of all other owner parameters $o_{1\ldots n}$ in the ownership tree. It is not immediately obvious why this condition is imposed --- it is dictated by features interacting with subtyping~\cite{ojcondition}.

A language designer could thus be posed the question, ``What are the consequences of this extra check? Specifically, does adding the check reject any correct programs that were accepted prior to adoption of this rule?" We answer this question with \toolname. First, we create a version of Ownership Java (OJ) called \emph{Reduced Ownership Java (ROJ)} where subtyping is prohibited and the additional constraint on owners is removed. As expected, \toolname\ fails to find a soundness counterexample in the ROJ language.

Now we are ready to query for the program of interest.  This program must (1)~fail the OJ typechecker,  (2)~pass the ROJ typechecker, and (3)~succeed at runtime.  This program is rejected by OJ, does not use subtyping, and is correct, per the three conditions in the query.  \toolname\ produces the following counterexample in just over a minute.
\begin{quote}
\footnotesize
\begin{verbatim}
class Main<O1, O2> {
    Main<world, O2> main(Main<O1, O2> arg) {
        return new Main<world, O2>();
    } }
new Main<world, world>.main(new Main<world, world>)
\end{verbatim}
\end{quote}
This program fails the Ownership Java typechecker because \texttt{world} is not a descendant of \texttt{O2} in the ownership tree. However, this program is otherwise correct: it passes our Reduced Ownership Java typechecker and does not fail at runtime. While the counterexample itself is not particularly profound, it provides evidence that suggests that a less restrictive static type rule could be designed.

The comparison between OJ and ROJ is made possible by the novel way in which \toolname\ allows users to compose sets of constraints when querying the solver. Crucially, by comparing two versions of a typechecker, we can better understand how the difference in type rules affects which programs are rejected.

\paragraph{Simplifying Counterexamples with Minimization}

Ownership Java programs can easily grow extremely large, due to the presence of unused classes and methods. In order to have an easily-understandable counterexample, we present our SMT solver with an additional goal: to minimize the size of the generated program. Thus, the counterexamples listed above are \emph{minimal:} no smaller program satisfies the constraints we imposed.

%     _       _   
%  __| | ___ | |_ 
% / _` |/ _ \| __|
%| (_| | (_) | |_ 
% \__,_|\___/ \__|
%                 

\subsection{Dependent Object Types (DOT)}

\hyphenation{Nano-DOT}
\hyphenation{Nano-Scala}

\emph{Dependent Object Types}, or DOT~\cite{dot} is both a formalization of the essence of Scala~\cite{scala-lang} and the basis for \emph{dotty}~\cite{dottycompiler}, an experimental Scala compiler. This makes DOT a good candidate for a real-world type system that underlies a common, modern, and practical programming language.

DOT features \emph{path-dependent types}. For example, the signature
\begin{quote}
\footnotesize
\begin{verbatim}
feed(pet : Animal, food: pet.FoodType) : Boolean
\end{verbatim}
\end{quote}
describes a function whose second argument has a type \texttt{pet.FoodType} that \emph{depends} on the first argument \texttt{pet}, hence the name path-dependent types.
DOT also provides record types such as \texttt{\{~name~:~String~\}}, and intersection types such as \texttt{type~Nat~=~Int~$\wedge$~Positive}.

In this section, we examine the Scala soundness issue SI-9633~\cite{amintatepaper, rosstatebug}.  The issue was discovered by reasoning about extending DOT with some features of full Scala, namely the addition of the \texttt{null} object. In this section, we demonstrate how to add \texttt{null} to DOT and use the \toolname\ Checker to find a counterexample to the issue\footnote{Our \toolname\ models of NanoDOT and NanoScala are available at \url{https://bitbucket.org/bonsai-checker/dot/}.}.

With the \toolname\ Checker library, we need just over 400~lines of code to implement DOT. The implementation closely follows the specification by~\citet{dotlanguage}. For example, three of the subtyping rules are below:
\begin{mathpar}
\inferrule{\Gamma \vdash T <: \top}{}
\and
\inferrule{\Gamma \vdash \bot <: T}{}
\and
\inferrule{\Gamma \vdash S <: T \\ \Gamma \vdash S <: U}{\Gamma \vdash S <: T \wedge U}
\end{mathpar}
Their corresponding implementations in \toolname\ are almost direct translations:
\begin{quote}
\footnotesize
\begin{verbatim}
(define/rec (dot-subtype? sub sup)
  (tree-match `(,sub . ,sup)
    '(_ . Any)             (lambda (T) #t)
     
    '(Nothing . _)         (lambda (T) #t)

    '(_ . (and . (_ . _))) (lambda (S T U)
                             (and (dot-subtype? sub T)
                                  (dot-subtype? sub U)))
    ...))
\end{verbatim}
\end{quote}

\paragraph{The NanoDOT language}

\begin{figure}[t]
\small
\begin{tabular}{p{0.5\columnwidth}p{0.5\columnwidth}}
\begin{grammar}
<term> ::= <term> "." <name> \hfill selection
\alt <definition>                                   \hfill object instantiation
\alt <name>                                         \hfill variable reference
\alt $\lambda$ "(" <name> ":" <type> ")" ":" <type> "{" <term> "}" \hfill lambda
\alt <term> "(" <term> ")"                          \hfill application

<definition> ::= "{" "val" <name> "=" <term> "}"    \hfill field
\alt "{" "type" <name> "=" <type> "}"               \hfill type
\alt <definition> $\wedge$ <definition>             \hfill aggregate
\end{grammar}
&
\begin{grammar}
<type> ::= $\top$ | $\bot$                         \hfill top, bottom
\alt <term> "." <name>                             \hfill type projection 
\alt "{" "val" <name> ":" <type> "}"               \hfill field decl 
\alt "{" "type" <name> ">:" <type> "<:" <type> "}" \hfill type decl 
\alt <type> $\wedge$ <type>                        \hfill intersection
\alt <type> $\rightarrow$ <type>                   \hfill function

<name> ::= "a" | "b" | "c" | $\varphi$
\end{grammar}
\end{tabular}
\caption{The NanoDOT language.}
\label{fig:nanodotgrammar}
\end{figure}

Figure~\ref{fig:nanodotgrammar} shows the syntax of the version of DOT we use to investigate the issue SI-9633. There are two deviations: first, for convenience, we add a global variable $\varphi$ of type $\top$ to the environment; its value will populate records in counterexamples. This addition makes counterexamples more readable: since DOT itself does not provide any primitive types, in pure DOT, \toolname\ would have to use type-tags as leaf values in  counterexamples. Second, we omit recursive types, because DOT's definition of recursive types using call-by-name semantics is incompatible with our call-by-value implementation.

We now discuss two counterexamples found using \toolname. The first explains a restriction imposed by DOT, while the second is a variant on the counterexample in SI-9633.

\paragraph{Disjoint domains in intersection types}

DOT restricts intersection types and aggregate definitions to field-disjoint terms. 
That is, {\sc AndDef-I} reads~\cite{dotlanguage}:
\begin{mathpar}
\inferrule{\Gamma \vdash d_1 : T_1 \\ d_2 : T_2 \\\\ \text{dom}(d_1) \cap \text{dom}(d_2) = \phi}{\Gamma \vdash d_1 \wedge d_2 : T_1 \wedge T_2}
\end{mathpar}
where $d_i$ are definitions and $\text{dom}(d)$ is the set of fields bound in $d$.

A reader lacking expertise in type systems might not immediately see the rationale for this restriction. 
Such a reader can simply remove {\sc AndDef-I} from our implementation of NanoDOT and observe the results. 
\toolname\ constructs the following useful counterexample:
\begin{quote}
\footnotesize
\begin{lstlisting}
$\lambda$ ( b: { val a: { val a: $\top$ } } ): $\top$ {
    b.a.a
} ( { val a = $\varphi$ } $\wedge$ { val a = { val a = $\varphi$ }} )
\end{lstlisting}
\end{quote}
When given this program, the interpreter crashes, finding that \texttt{b.a} is undefined. 
Since \texttt{b} is bound to an intersection where both sides have a value for the field \texttt{a}, the value \texttt{b.a} could be either $\varphi$ or \texttt{ \{~val~a = $\varphi$~\}}.
If the left side is chosen, we get a runtime error. 
If the right side is chosen, we get $\varphi$. 
Since our implementation gives precedence to the left member of an intersection, our interpreter throws a runtime error, making this program a counterexample. 
In summary, this example points out that if the interpreter is to make a greedy choice among intersected values, these values must have no conflicting fields, which explains {\sc AndDef-I}.

\paragraph{Collapsing the subtype hierarchy using bad bounds (SI-9633)}

A common soundness issue with dependent types relates to \emph{bad bounds}, which allows creating uninhabitable types. For instance, the type \texttt{ \{~type S >: $\top$ <: $\bot$~\}}, which means that $\top$ \texttt{<:} $S$ \texttt{<:} $\bot$, should clearly be uninhabitable. Otherwise, we would be able to collapse the subtype hierarchy and prove all types equivalent, much like how a single contradiction allows proving anything in an inconsistent logical system~\cite{dotlanguage}.

DOT does not check the bounds of dependent \emph{types} but ensures that \emph{values} have types with correct bounds. So, one can create the badly-bounded type \texttt{\{~type~S >:~$\top$ <:~$\bot$~\}} but values are syntactically constrained to have coinciding lower and upper bounds, using the construct 
\texttt{\{~type~S =~$\top$~\}}, which expands to \texttt{\{~type~S >:~$\top$ <:~$\top$~\}}.
As a result, while \emph{types} with bad bounds can be created in DOT, it is impossible to instantiate \emph{values} with badly-bounded dependent types.  This restriction makes it easy to check the bounds of a value upon its instantiation. 

In Scala, however, it is possible to create a value without going through a constructor, which makes bounds checking hard.  
\citet{oderskypost} describes three situations where this is possible: using \texttt{lazy} for delayed instantiation, using \texttt{null} for uninstantiated values, and using type projection. In these cases, the compiler cannot check for bad bounds of dependent types. In Scala, it \emph{is} possible to instantiate values with badly-bounded dependent types. This is the basis for SI-9633.

Scala soundness bugs, including SI-9633, are traditionally demonstrated with an example that casts an arbitrary value to the type \texttt{Nothing}. Since \texttt{Nothing} is a subtype of all other types, this allows us to cast a value to any other value, for example, via the sequence \texttt{Number} $\rightarrow$ \texttt{Nothing} $\rightarrow$ \texttt{String}. 
In practice, such an operation leads to the runtime \texttt{ClassCastException} in the JVM.

To explore bugs related to bad bounds, we extend NanoDOT to \emph{NanoScala} with a \texttt{null} value of type $\bot$. We also add a term to allow casting.
The term \texttt{cast $t$ to $T$} is typechecked in the usual manner, by ensuring that $t : S \wedge S\; \texttt{<:} T$.
At runtime, the \texttt{cast} performs the same check to detect counterexample programs.  This is sufficient to model the relevant Scala issues in NanoScala.

We are now ready to query \toolname\ for a NanoScala equivalent of the SI-9633 counterexample.  This program was found in around 4 minutes:
\begin{quote}
\footnotesize
\begin{lstlisting}
cast
    ($\lambda$ (a: {type b >: $\top$ <: $\bot$}): a.b {
        $\varphi$
    })(cast null to {type b >: $\top$ <: $\bot$})
to $\bot$
\end{lstlisting}
\end{quote}
Why does this program typecheck?  
The inner cast typechecks because \texttt{null} is of type $\bot$, which is a subtype of all types --- this includes the type \texttt{\{~type~S >:~$\top$ <:~$\bot$~\}}, regardless of its bad bounds. 
The outer cast to $\bot$ typechecks because the type \texttt{a.b} is bounded above by $\bot$; this is allowed because badly-bounded types can be created in DOT.

Why does the program fail at runtime?
At runtime, the issue occurs when \texttt{null} is cast to \texttt{\{~type~S >:~$\top$ <:~$\bot$~\}}.  This type is uninhabited, yet, modeling Scala, NanoScala nevertheless casts \texttt{null} to it. The execution continues with a value with badly-bounded dependent types. 
As a consequence, we can now invoke the lambda expression synthesized by \toolname.  This lambda should not be called, since its argument is of an uninhabited type. However, when called anyway, it returns $\varphi : \top$ which is cast to $\bot$. This raises an exception.

In summary, NanoScala allows us to cast a non-\texttt{null} value to $\bot$, which is at the heart of the SI-9633 counterexample. This is because the NanoScala \texttt{null} and \texttt{cast} terms violate NanoDOT's property that all dependent types are backed with an instantiated value.

There are two important notes regarding our implementation of DOT:
First, to prevent \toolname\ from synthesizing counterexamples related to null dereferencing, we must tell \toolname\ to disregard bugs involving the selection operator. 
Second, our current implementation of NanoDOT does not maintain type information at runtime. Thus, we only raise an exception on casts of non-\texttt{null} values to $\bot$, which can be detected without any runtime type information. This is sufficient to catch the runtime errors we are interested in; the Scala example \texttt{Number} $\rightarrow$ \texttt{String} would fail when \texttt{Number} is cast to \texttt{Nothing}, which we model as DOT $\bot$.

Finally, we can translate the NanoScala counterexample to Scala. We must manually make two minor changes.
First, unlike DOT, Scala does check type annotations that have bad bounds. A simple layer of indirection suffices to avoid this check: we create our badly-bounded type \texttt{\{~type~S >:~$\top$ <:~$\bot$~\}} as an intersection of two well-bounded types \texttt{P} and \texttt{Q}:
\begin{quote}
\footnotesize
\begin{verbatim}
trait P {type B <: Nothing}
trait Q {type B >: Any}
\end{verbatim}
\end{quote}
Note that Scala's intersection types are not commutative due to shadowing, so we must care to use either \texttt{P with Q} or \texttt{Q with P}, depending on the context.

The second change is purely syntactic. Scala's lambdas do not allow path-dependent return type annotations in the same style as methods, so we translate the lambda to a named method.
\begin{quote}
\footnotesize
\begin{verbatim}
def bad(a: P with Q): a.B = 123
bad(null: Q with P): Nothing
\end{verbatim}
\end{quote}
When run with \texttt{scala}, this example produces:
\begin{quote}
\footnotesize
\begin{verbatim}
java.lang.ClassCastException:
java.lang.Integer cannot be cast
to scala.runtime.Nothing
\end{verbatim}
\end{quote}
After variable renaming, this counterexample corresponds almost exactly with the one reported on SI-9633. Figure~\ref{fig:si-9633} presents a side-by-side comparison. 

\begin{figure}
\begin{tabular}{p{5cm}p{5cm}}
\footnotesize
\begin{verbatim}
trait A { type L <: Nothing }
trait B { type L >: Any }
def toL(b: A with B): b.L = 123
val p: B with A = null
toL(p): Nothing
\end{verbatim}
&
\footnotesize
\begin{verbatim}
trait A { type L <: Nothing }
trait B { type L >: Any }
def toL(b: B)(x: Any): b.L = x
val p: B with A = null
println(toL(p)("hello"): Nothing)
\end{verbatim}
\end{tabular}

\caption{Comparing \toolname's counterexample with SI-9633. 
Left: A counterexample found by \toolname, manually translated to Scala with variables renamed for clarity.
Right: The counterexample reported on SI-9633~\cite{rosstatebug}, copied verbatim.
}
\label{fig:si-9633}
\end{figure}

\paragraph{Discussion}

We demonstrate the use of \toolname\ in studying DOT, a language with a complex type system. We provide two subtle bugs that can be introduced in DOT by modifying the language, and show that in each case \toolname\ produces elegant counterexamples after only a couple of minutes. The latter of these counterexamples reproduces a soundness counterexample for the Scala compiler, SI-9633.

%%%%%%%%%%%%%%%%%%%%%%%%%%%%%%%%%%%%%%%%%%%
\section{Empirical Evaluation of Performance}
\label{sec:performance}

In this section, we evaluate the efficiency of \toolname. We consider the speed of symbolic evaluation and solving, the sizes of the counterexamples, and the sizes of program spaces that are explored.
We find that \toolname\ reliably finds bugs in a few seconds or minutes.
When left to run for longer periods of time, \toolname\ explores spaces containing programs much larger than the minimal counterexample, providing a margin of assurance.

\subsection{Comparing \toolname\ with Fuzzers}
\label{sec:eval:scalabilityandfuzzing}

Here, we compare \toolname\ to the fuzzers mentioned in Section~\ref{sec:relatedwork}: in particular, the syntactic fuzzer built into Redex~\cite{redexfuzzer}, and the type fuzzer based on judgement trees~\cite{makingrandomjudgements}.

\paragraph{The benchmark}

We implemented\footnote{Our \toolname\ model of stlc+lists, with patchfiles to introduce each bug, is available at \url{https://bitbucket.org/bonsai-checker/stlc-benchmark/}.}
the ``stlc+lists" language from the Redex benchmark~\cite{redexman} in \toolname\ and introduced each of the nine bugs to our implementation (these were all one-line changes). We compared the performance of \toolname\ to the performance of the syntactic fuzzer and type fuzzer from the Redex benchmark in two respects: first, the time taken to find a counterexample, and second, the average size of counterexamples found. Note that all tests were run on the same machine.

\paragraph{Results}

Our results are plotted in Figure~\ref{fig:graph1} (left), and the times and sizes, correlated by bug, are listed in Table~\ref{tab:fuzzers}. We make two important observations below.

First, we note \toolname's consistency. While fuzzers are slightly faster than \toolname\ on some benchmarks, they are several orders of magnitude slower on others, taking anywhere from a few minutes to over an hour. \toolname, on the other hand, takes between 1 and 5 seconds on every benchmark. Consider, for example, bugs 4 and 5, which were the hardest for fuzzers to find, despite both being labeled ``shallow" errors in the benchmark suite. Bug 5 changes the interpreter to return the head of a list when \texttt{tail} is applied, while bug 4 assigns the return type of \texttt{cons} to \texttt{int}, rather than \texttt{Listof int} as expected. \toolname\ finds these bugs just as fast as it finds other bugs. However, these bugs are many orders of magnitude more difficult for fuzzers to reveal. This is because these bugs require a specific set of interactions in the counterexample program, and thus the probability of randomly generating such a program is extremely low.

Second, we compare the sizes of counterexamples produced by the two tools. The type fuzzer, though efficient, generates extremely large and complex programs as counterexamples. This likely contributes to its effectiveness, since larger programs are likelier to uncover soundness bugs~\cite{csmith}. However, such programs are extremely difficult for users to reason about, with (for example) dozens of layers of nested lambdas.
In contrast, \toolname\ consistently produces small counterexamples. In almost all cases, \toolname's counterexamples were identical to the ones suggested by the Redex benchmark authors --- one was smaller.

Inspired by the latter observation, we evaluated \toolname's efficiency on constructing \emph{minimal} counterexamples by querying the solver for a solution that minimizes the size of the counterexample. Our results are shown in Table~\ref{tab:minimize}. We find that the most of the counterexamples produced by \toolname\ were already minimal, even without the minimization query. However, for the three bugs that could be further minimized, the minimization query only took 10\%-30\% longer than the standard query.

As a final demonstration of \toolname's efficiency compared to fuzzers, we consider a historic bug involving let-polymorphism in the presence of mutable references, in e.g. ML~\cite{mlbug1, mlbug2}. This ``classic let+references" bug is included in the Redex benchmark as let-poly-2. However, the syntactic fuzzer has been run on it for several days with no results. The type fuzzer cannot model let-poly due to the presence of polymorphism, which requires the generator to make parallel choices that must match up, and CPS-transformed type judgements, which impede its termination heuristics~\cite{makingrandomjudgements}. \toolname, on the other hand, finds this bug in just over twenty minutes\footnote{Our \toolname\ model of let-poly-2 is available at \url{https://bitbucket.org/bonsai-checker/let-poly/}.}, with a counterexample almost identical to the one presented in literature~\cite{mlbug2}.

\subsection{Comparing \toolname\ with Pipal}
\label{sec:eval:korat}

We compare the scalability of \toolname\ to the the scalability of Pipal as described by \citet{roberson}. This comparison is only an approximation because \toolname\ and Pipal encode different search spaces (symbolic programs and symbolic intermediate states, respectively). Thus, for example, \toolname\ must model \emph{all} heap objects, while Pipal must limit the exploration to four heap objects and $n$ integer literals. Furthermore, the Pipal experiments were performed using different symbolic execution frameworks and solvers. Nevertheless, Pipal results for comparable trials are listed in Table~\ref{tab:langs}.

These measurements illustrate that \toolname\ and Pipal generally scale to around the same order of magnitude. Thus, we find that even though \toolname\ performs symbolic execution on both the typechecker and the interpreter, it still has comparable performance to Pipal, which only performs symbolic execution on the typechecker. That is, compared to Pipal, \toolname\ does not sacrifice efficiency for its versatility and ease-of-use.

\begin{table*}
{\small
\begin{tabular}{l |
S[table-format=4.3, table-align-text-post=false]
S[table-format=4.3, table-align-text-post=false]
S[table-format=4.3, table-align-text-post=false]
|
S[table-format=2.1, table-align-text-post=false]
S[table-format=2.1, table-align-text-post=false]
S[table-format=2.1, table-align-text-post=false]
}
\hline
Bug &
\multicolumn{1}{c}{Bonsai (sec)} &
\multicolumn{1}{c}{Syntax/Bonsai} &
\multicolumn{1}{c}{Type/Bonsai} &
\multicolumn{1}{c}{Bonsai} &
\multicolumn{1}{c}{Syntax/Bonsai} &
\multicolumn{1}{c}{Type/Bonsai} \\
\hline
 &   & \multicolumn{1}{c}{Time} &   &   & \multicolumn{1}{c}{Size} &   \\
\hline
stlc-1 & 0.86{s} & 0.61 & 1 & 5{n} & 1 & 57 \\
stlc-2 & 0.99{s} & 56 & 7.7 & 9{n} & 1 & 14 \\
stlc-3 & 0.84{s} & 0.25 & 0.054 & 5{n} & 3.6 & 51 \\
stlc-4 & 3{s} & 1200{+} & 34 & 17{n} & {N/A} & 1.8 \\
stlc-5 & 0.99{s} & 3600{+} & 40 & 9{n} & {N/A} & 14 \\
stlc-6 & 2{s} & 1000 & 15 & 17{n} & 0.76 & 5.7 \\
stlc-7 & 1{s} & 8 & 0.07 & 9{n} & 2 & 19 \\
stlc-8 & 4.6{s} & 14 & 0.061 & 29{n} & 0.88 & 6.2 \\
stlc-9 & 1.2{s} & 0.4 & 0.049 & 15{n} & 1.4 & 13 \\
\hline
Full suite & 16{s} & 600{+} & 12 & & & \\
\hline
let-poly-2 & 1200{s} & {24-hr timeout} & {Cannot find} & 47{n} & {N/A} & {N/A} \\
\hline
\end{tabular}
}
\vspace{0.2cm}

\caption{The performance of \toolname\ and the syntax and type fuzzers described in the text. Note that fuzzer statistics are provided as ratios against \toolname\ statistics. ``+" indicates a lower bound due to a one-hour timeout. Size is measured in nodes (n) as per the Redex benchmark: ``the number of pairs of parentheses and atoms in the s-expression representation of the term"~\cite{redexman}. Figure~\ref{fig:graph1} (left) represents this data on a scatter plot. Note that the type fuzzer cannot model let-poly-2 due to polymorphism and CPS-transformed judgement rules~\cite{makingrandomjudgements}. }
\label{tab:fuzzers}
\end{table*}

\begin{table*}
{\small
\begin{tabular}{l |
S[table-format=1.2, table-align-text-post=false]
S[table-format=2.1, table-align-text-post=false]
}
\hline
Bug &
\multicolumn{1}{c}{Time (sec)} &
\multicolumn{1}{c}{Min. Size} \\
\hline
stlc-1 & 0.92{s} & 5{n} \\
stlc-2 & 1.1{s}  & 9{n} \\
stlc-3 & 0.99{s} & 5{n} \\
stlc-4 & 2.7{s}  & 17{n} \\
stlc-5 & 1.0{s}  & 9{n} \\
stlc-6 & 2.6{s}  & 13{n*} \\
stlc-7 & 1.2{s}  & 5{n*} \\
stlc-8 & 4.9{s}  & 21{n*} \\
stlc-9 & 1.3{s}  & 15{n} \\
\hline
\end{tabular}
}
\vspace{0.2cm}
\caption{The performance of \toolname\ on constructing minimal counterexamples. The star (*) indicates that the minimized counterexample is smaller than the one produced without the minimization query.}
\label{tab:minimize}
\end{table*}

\subsection{The \toolname\ Encoding}
\label{sec:eval:scalability}

Here, we compare the classical ``syntax tree" described in Section~\ref{sec:bonsai-trees} (with subtree sharing) against the \toolname\ tree. We evaluate the time required to check a type system as a function of the program space size. We break down the time into symbolic evaluation and solving. We find that the \toolname\ tree scales significantly better than the classical syntax tree, allowing us to explore much larger search spaces in the same amount of time.

\begin{figure}[t]
\begin{tabular}{r  l}
\includegraphics[width=0.58\linewidth]{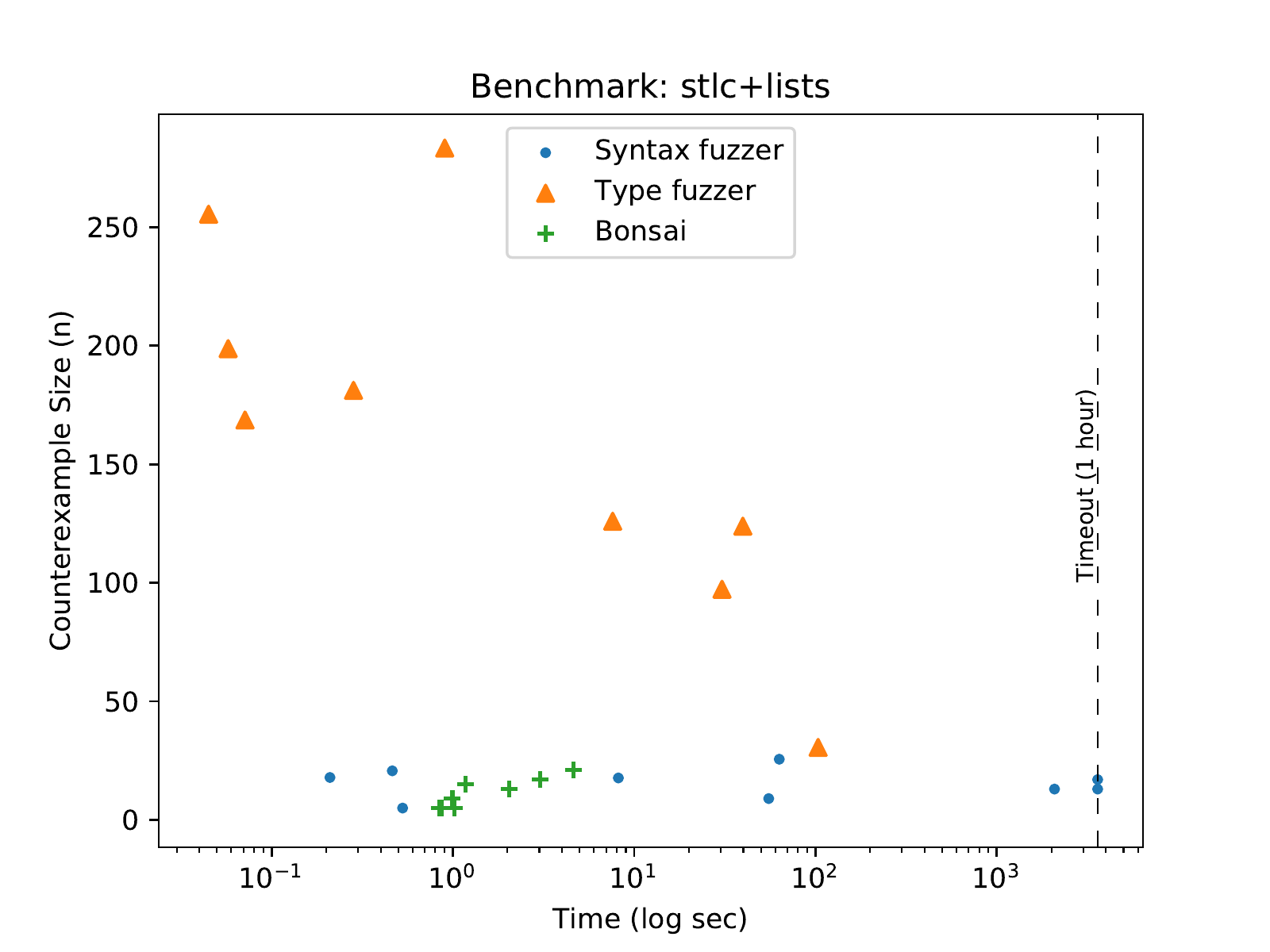} &
\includegraphics[width=0.40\linewidth]{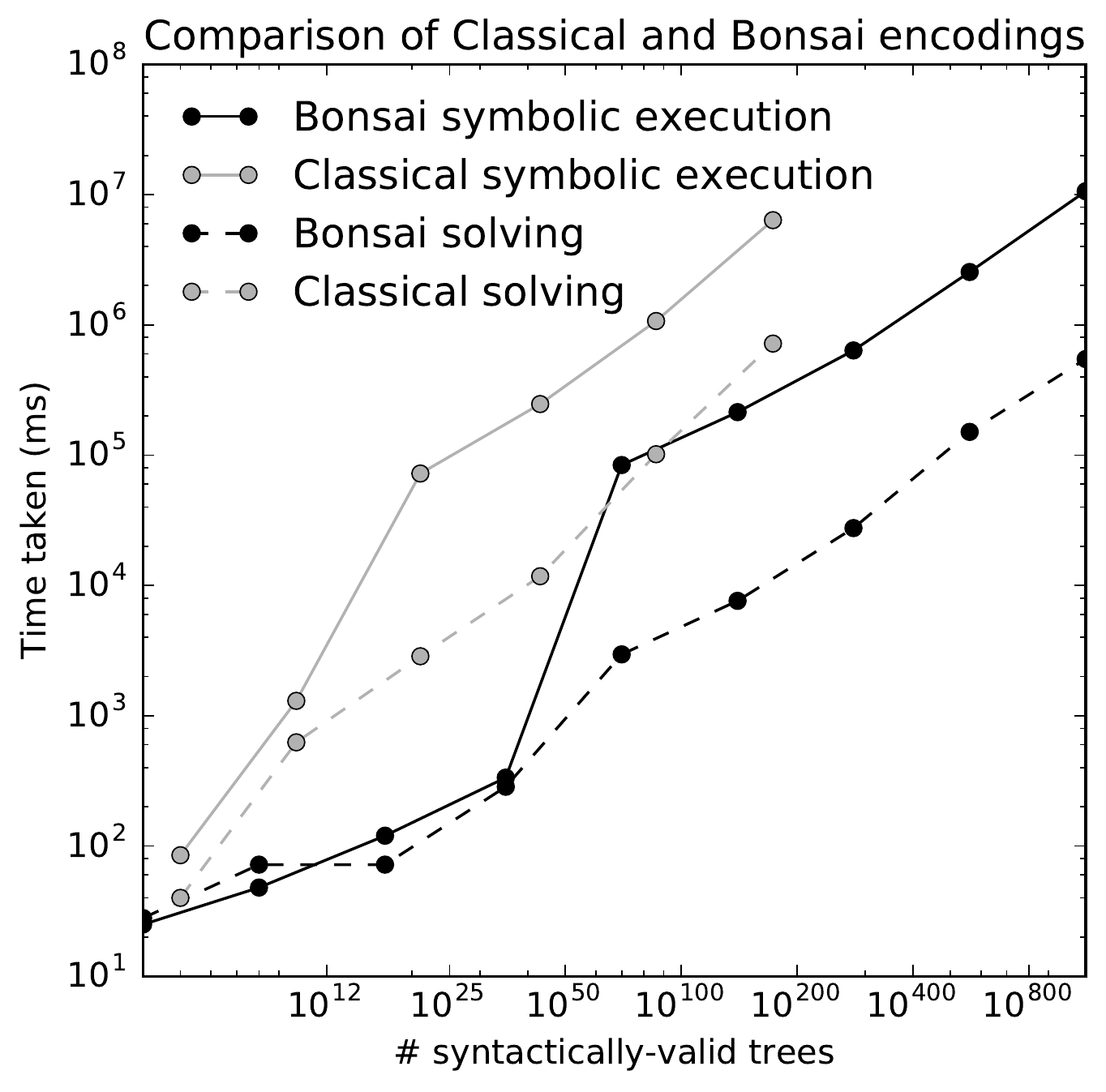} \\
\end{tabular}
\caption{Left: Comparing fuzzers and \toolname\ on the Redex benchmark.
Right: Comparing the classical syntax tree and \toolname\ on verifying the lambda calculus.}
\label{fig:graph1}
\end{figure}

\paragraph{The benchmark}

We implemented a simple recursive-descent interpreter for the lambda calculus, bounded at 4 recursive calls. We then added a `typechecker' that checks for free variables. We searched for counterexamples using both encodings, with the \emph{same} typechecker and interpreter. Thus, the only variable between the trials is the symbolic tree encoding used and its size.

In this setup, there are no possible counterexamples. This is intentional: a satisfiable formula requires the solver to reach \emph{any} valid model, whereas an unsatisfiable formula requires the solver to visit \emph{all} models and prove that none are satisfiable. Thus, an unsatisfiable formula represents a more taxing stress test for the solver. In practice, we find that satisfiable queries almost always take less than half the time that unsatisfiable queries take, and are usually even faster.

Since tree `size' has different meanings for the classical and \toolname\ encodings, we normalize by the number of \emph{syntactically valid} ASTs represented by a symbolic tree of a given depth.

\paragraph{Results}

The graph in Figure~\ref{fig:graph1} (right) shows that the \toolname\ encoding allows us to scale to a much larger number of ASTs than the classical encoding in the same amount of time. For example, if we impose a time limit of approximately 20 seconds, then the classical encoding explores less than $10^{25}$ ASTs while the \toolname\ encoding explores more than $10^{50}$ ASTs. Similarly, if we want to explore $10^{25}$ trees, then \toolname\ is roughly a thousand times faster than the classical encoding. In our experience, the classical encoding does not scale to be able to catch most of the counterexamples discussed in this paper.

Note that the \toolname\ encoding consistently spends less time in the solver than the classical encoding, even though \toolname\ pushes the complexity of the formal grammar to the solver. This suggests that even though \toolname\ often creates larger symbolic trees than the classical encoding to represent the same AST, it has a performance advantage in both symbolic execution \emph{and} the solver.

In practice, we rarely search ASTs beyond depth 10; smaller trees are always enough to synthesize counterexamples identified by experts (see Table~\ref{tab:langs}). Nevertheless, we provide data for much larger trees in this graph to show that the encoding is scalable to much larger spaces if needed.

\subsection{Summary of Case Studies}

\paragraph{The benchmark}

We implemented several different languages with \toolname, and searched for counterexamples by planting bugs in the typecheckers. We collected data for the smallest parameters needed to catch these bugs, as well as for scaled-up trials. The scaled-up trials were run with the same queries, but parameters modified in order to search larger spaces.

\paragraph{Results}

Table~\ref{tab:langs} lists data collected for several languages implemented with \toolname. We find that each of the counterexamples, including those found by experts, can be synthesized by \toolname\ within a few minutes. This suggests that \toolname\ is suitable for interactive use while developing a type system.

Additionally, we find that \toolname\ scales to much larger search spaces effectively, searching through larger, more complex programs. Indeed, the new `scaled-up' counterexamples were often several times as large as the `minimal' counterexamples. Thus, while bounded model checking is not a substitute for a formal proof, these results suggest that \toolname\ can be used to search through large spaces of programs efficiently to provide confidence that the type system is most likely sound---or, alternatively, reveal a large counterexample that may not be found by a human.

Finally, we note that \toolname\ performs well on a wide variety of languages, allowing users to easily implement features such as closures, environments, classes/objects, ownership types, dependent types, records, and polymorphism in their languages.

\begin{table*}[t]
    \centering
    {\small
    \begin{tabular}{ p{2cm}  p{2cm}  p{2cm}  r  r  r  r  | r  r  r }
        \hline
        Language
        & Description
        & Parameters
        & $|S|$
        & $|ce|$
        & $t_\text{\toolname}$
        & $t_\text{Pipal}$
        & $|S_\text{+}|$
        & $|ce_\text{+}|$
        & $t_\text{+}$
        \\
        \hline

        Arithmetic
        & {\raggedright Typed arithmetic~\cite{piercebook}}
        & depth = 9\newline
        no bound
        & $10^4$
        & 13
        & 1s
        & 0.5s
        & $10^{88}$
        & 99
        & 11m \\
        
        Simply-typed $\lambda$-calculus
        & {\raggedright From Redex~\cite{redexman}, bug 5}
        & depth = 5\newline
        bound = 4
        & $10^4$
        & 9
        & 0.97s
        & ---
        & $10^{21}$
        & 43
        & 1.7m \\
        
        \hline
        
        Featherweight Java
        & {\raggedright A minimal core calculus for Java with inheritance~\cite{featherweightjava}}
        & \# classes = 3\newline
        \# methods = 3\newline
        depth = 6\newline
        bound = 7 % 6+1
        & $10^{27}$
        & 42
        & 3s
        & 2.1s
        & $10^{119}$
        & 163
        & 13m
        \\
        
        Ownership\newline Java
        & {\raggedright Featherweight Java augmented with ownership types~\cite{ownershipjava}}
        & \# classes = 2\newline
        \# methods = 4 \newline
        \# owners = 2 \newline
        depth = 8 \newline
        bound = 7 % 6+1
        & $10^{63}$ 
        & 72
        & 17s
        & $>$250s
        & $10^{342}$
        & 158
        & 30m
        \\
        
        \hline
        
        NanoDOT
        & {\raggedright Subset of dependent object type calculus~\cite{dotlanguage}}
        & depth = 10\newline
        obj depth = 6\newline % 3 + 1
        no bound
        & $10^{19}$
        & 23
        & 211s
        & ---
        & $10^{39}$
        & 49
        & 8m
        \\
        
        NanoScala
        & {\raggedright NanoDOT augmented with casts and \texttt{null}}
        & Same as NanoDOT
        & $10^{20}$
        & 24
        & 175s
        & ---
        & $10^{64}$
        & 39
        & 48m
        \\
        \hline
    \end{tabular}
    }
    \nocaptionrule
    \caption{
    An overview of the languages implemented with \toolname. Here, $t_{\toolname}$ is the time required to catch the soundness bug described in the text, $|ce|$ is the size of the counterexample measured as in Table~\ref{tab:fuzzers}, and $|S|$ is the number of syntactically-correct programs explored. $t_\text{+}$, $|S_\text{+}|$, and $|ce_\text{+}|$ are analogous, but for long-running trials. We show relevant comparisons with Pipal where possible.
    }
    \label{tab:langs}
\end{table*}

\section{Related Work}
\label{sec:relatedwork}

Most existing tools for type system designers are focused on checking for soundness bugs. Here, we discuss the various techniques used to detect such bugs. Empirical evaluations of \toolname\ against these tools are presented in Section~\ref{sec:performance}. Additionally, we discuss additional related work in theorem-proving and symbolic execution.

\subsection{Fuzzing}

A type system fuzzer generates random programs and tests them on the typechecker and interpreter until it finds one that (1)~passes the typechecker, and (2)~fails in the interpreter. %Ideally, no such programs would exist. So, 
If such a program is found, it witnesses a soundness bug.

Fuzzing presents serious scalability challenges. Recall Figure~\ref{fig:pipeline}, which shows a high-level overview of the fuzzing process. At each step, the parser, typechecker, or interpreter rejects the vast majority of its inputs, making the probability of a random program witnessing a soundness error very small. Modern fuzzers address this problem by carefully generating random programs that are likelier to be counterexamples. For example, \emph{syntactic fuzzers} such as the one used by PLT Redex~\cite{redexfuzzer} generate random syntactically-valid programs, bypassing the parser. 
%Each program generated by a syntactic fuzzer is thus much likelier to be a counterexample.
%
Syntactic fuzzers have been shown to find a wide variety of bugs in real languages~\cite{redex}. However, they are still not scalable enough to catch many simple bugs~\cite{makingrandomjudgements}. 

A \emph{type fuzzer} generates random well-typed programs, bypassing both the parser and the typechecker. Type fuzzing can be done in many ways: for example, by using constraint logic programming to express the typechecker declaratively (this has been used to fuzz Rust~\cite{rustfuzzing}) or by generating random type judgment trees and then using that to derive a program (this has been used to fuzz GHC~\cite{makingrandomjudgements}).

% \subsection{How useful are fuzzers?}

% (See \tabref{comparison} for a condensed summary of this section.)

%\paragraph{Strengths}
Fuzzing is an effective technique for two reasons. First, a fuzzer can easily check actual implementations of languages. This eliminates the need to formalize the language, and allows users to catch implementation-dependent bugs that would be missed by a formalization. Second, a fuzzer produces a concrete counterexample program, which makes it easy to diagnose and fix the soundness issue. This claim is supported by a recent study, which used NanoMaLy~\cite{nanomaly} to produce counterexamples that cause ill-typed ML programs to crash. Students who were given NanoMaLy's counterexamples in an exam setting were 10\% to 30\% more likely to correctly explain and fix the type error than students who were given only a printout of the compiler error.

%\paragraph{Weaknesses}
However, fuzzers have a critical weakness: by their nature, random fuzzers are non-exhaustive. If after 24 hours no soundness error has been discovered, we cannot assume that the typechecker is sound. Indeed, neither syntactic fuzzers nor type fuzzers are able to discover certain simple bugs in the PLT Redex benchmark~\cite{redex} after many hours~\cite{makingrandomjudgements}, because those bugs are only witnessed by a small set of programs that are extremely unlikely to be generated randomly.

\subsection{Handling non-exhaustiveness with Pipal}
One specialized kind of type fuzzer \emph{does} make some exhaustiveness guarantees. Pipal~\cite{pipal, roberson} requires a typechecker to be encoded as a set of constraints imposed on a finite set of intermediate program states. At each iteration, Pipal queries a constraint solver for a well-typed state and performs a single step of evaluation to see whether \emph{progress} or \emph{preservation} were violated.

In order to search the space of intermediate states efficiently, Pipal carefully monitors how the intermediate state was manipulated during the single step of evaluation. If the program is not a counterexample, Pipal uses this information to add more constraints to the set of programs, effectively pruning the search space for the next iteration. Eventually, if no counterexample is found, the solver returns UNSAT, which implies that the search space has been exhaustively checked. That is, no state within that space witnesses a soundness bug.

While Pipal is scalable, it requires typecheckers to be manually rewritten as a set of declarative constraints and carefully-formulated invariants on intermediate states. Not only is this tedious for the user, but it also prevents users from directly checking the \emph{implementations} of typecheckers. Thus, Pipal loses one of the primary benefits of fuzzing, which is to check executable implementations directly.

The \toolname\ algorithm, on the other hand, combines the ease-of-use of a fuzzer with the scalability of Pipal. \toolname\ uses symbolic execution to \emph{automatically} convert an executable language implementation into constraints for a solver, and it \emph{directly} solves for a counterexample with a single call to the solver. In this sense, \toolname\ may be regarded as a final successor to the type fuzzer, deriving many of Pipal's scalability benefits ``for free" without any additional effort on the part of the user. Importantly, \toolname\ searches for full executable programs as counterexamples: these programs are easier to understand than the intermediate states reported by Pipal. Finally, unlike Pipal's iterative algorithm, \toolname\ makes only one query to the solver. This property allows us to ask the solver a variety of questions beyond soundness, and is at the heart of \toolname's versatility.

\subsection{Theorem Proving}

%

% Ideally, we would be able to mathematically prove the soundness of type systems used by all popular programming languages.
If a fuzzer or a bounded model checker fails to find a witness to a soundness bug, then we cannot claim to have a proof of soundness: there is always the possibility that a program larger than the bound might expose a bug. In practice, the small-scope hypothesis~\cite{smallscope} conjectures that all soundness bugs will be revealed by small witness programs, and thus for a sufficiently large search bound the failure to find a counterexample is usually compelling evidence (but not proof) that the type system is correct.

Unlike fuzzing or bounded model checking, however, a mathematical proof of soundness provides \emph{complete} assurance that a type system is sound.
%
%A formal proof of soundness generally consists of three parts~\cite{piercebook}. First, we formalize the language in terms of syntax, typing rules, and evaluation rules. Next, we prove the \emph{progress} theorem: a well-typed program never gets `stuck'. Finally, we prove the \emph{preservation} theorem: a well-typed program remains well-typed after a reduction is applied. These two proofs are usually inductive on term size.
%
Such proofs can be developed manually or using proof assistants such as~\cite{coq} and~\cite{isabelleprover}. For instance, \citet{javaprobably} prove soundness of a subset of Java manually, while~\citet{javadefinitely} do so using Isabelle/HOL~\cite{isabelleprover}.

Unfortunately, theorem-proving requires a formalized model of the language, such as Featherweight Java~\cite{featherweightjava} for Java and DOT~\cite{dot} for Scala. Most modern languages are too complex to be fully formalized, and the models that do emerge often take years of development. Furthermore, the actual proofs are often long and tedious, requiring significant manual effort. For example, the mechanized proof of the DOT language~\cite{dot} using the Coq proof assistant~\cite{coq} consists of several thousand lines of code~\cite{dotcoq}, even though the language can be formalized in a couple of pages~\cite{dotlanguage}. We hope that \toolname\ may aid in model design in a fashion similar to PLT Redex~\cite{redex}.

% Proof assistants still require significant manual effort, because one must first encode the language in the proof assistant, and then guide the automated theorem prover towards a proof. For example, 
% We hope that \toolname\ can be extended to synthesize some artifacts of the proof, offering help in the manner that SMT solvers aid theorem provers~\cite{DBLP:conf/itp/BlanchetteN10}.
% Most modern languages are too complex to be fully formalized. So, attempts to prove them sound operate on featherweight models such as . These models take years to emerge and we hope that \toolname\ may aid in model design in a fashion similar to PLT Redex~\cite{redex}. 

%Furthermore, they do not necessarily resemble the actual implementation of the language. For instance, languages are rarely implemented purely through reduction relations. Thus, models may miss implementation bugs that are specific to a particular typechecking algorithm or interpretation technique. So, even though a formal proof is the gold standard of `correctness', real-world languages are rarely proven sound.

\subsection{Symbolic execution and synthesis}

Bounded verification tools~\cite{CBMC,JavaPathFinder,galeotti2010analysis,dolby2007finding,taghdiri2007automating,Dennis2006,Torlak:2010:MemSAT,sketching2,kaplan} encode the concrete semantics of the language and translate the program (or just one execution path) into a logical formula.  The formula represents constraints whose solution answers queries about symbolic inputs to the program.  Symbolic execution has become competitive in advanced analysis tasks, such as analysis of high-order contracts~\cite{DBLP:journals/corr/NguyenTH15}.

Generation of a counterexample program is related to program synthesis because we can think of the type system as the specification: the desired program passes the typechecker and fails in the interpreter.  Synthesizers can be roughly divided into three kinds: rewriting~\cite{FFTW05,spiral,autobayes},  deductive~\cite{pingali-bernoulli,specware,srivastava-popl10}, and those based on searching a space of programs with contraint solvers~\cite{DBLP:journals/csur/AngliunS83,DBLP:conf/popl/Gulwani11, sketching2,vechev-pldi06,srivastava-popl10}.  While the first two categories derive a program from a specification, the last category searches a space of programs, conceptually evaluating each against the specification. This search is analogous to \toolname's search for the counterexample program.

Reducing the path explosion during symbolic evaluation has been previously addressed in ESC/Java~\cite{DBLP:conf/popl/FlanaganS01} and Rosette~\cite{DBLP:conf/pldi/TorlakB14}. 

Bonsai trees are a so-called \emph{symbolic data structure}~\cite{DBLP:conf/snapl/BodikCPY17}: despite producing an unusual encoding when symbolically evaluated, they offer programmers the usual AST interface, facilitating specification of compact (big step) typecheckers and interpreters.

\section{Conclusion}
\label{sec:conclusion}

\toolname\ is a type system designer's assistant, which uses symbolic evaluation to convert typecheckers and interpreters into constraints. By combining these constraints, \toolname\ can query a constraint solver for programs with specific properties. Such queries can be used to find soundness errors (including intricate bugs that took experts many years to discover), compare two type systems, find unnecessary restrictions, and even synthesize simple typechecks automatically.

\toolname\ uses a novel data structure to encode sets of abstract syntax trees. This allows it to scale to extremely large search spaces, while still being fast enough to use interactively while designing a type system.
Together, these results suggest that \toolname\ can significantly aid type system designers in developing novel type systems.

{\raggedright
\bibliographystyle{abbrvnat}
\bibliography{references,bibliography,references-bodik}}

%\newpage
%\input{A-Appendix}

\end{document}